\documentclass[11pt,a4paper]{article}
\usepackage{jheppub}
\usepackage{amsmath,amsfonts,amssymb}
\usepackage{mathtools}
\usepackage{braket}
\usepackage{multirow}
\newcommand{\be}{\begin{equation}}
\newcommand{\ee}{\end{equation}}
\newcommand{\ba}{\begin{eqnarray}}
\newcommand{\ea}{\end{eqnarray}}
\def\bs{\begin{subequations}}
\def\es{\end{subequations}}

\def\vp{\varphi}

\def\cR{{\cal R}}

\def\p{\partial}

\newcommand{\bea}{\begin{eqnarray}}
\newcommand{\eea}{\end{eqnarray}}

\newcommand{\half}{\tfrac{1}{2}}

\newcommand{\Rb}{\bar{R}}

\newcommand{\bi}{\begin{itemize}}
\newcommand{\ei}{\end{itemize}}

\newcommand{\gb}{\bar{g}}
\newcommand{\eq}[1]{\begin{align} #1 \end{align}}
\newcommand{\spliteq}[1]{\begin{align}\begin{split}\begin{aligned} #1 \end{aligned}\end{split}\end{align}}

\newcommand{\del}{\partial}

\newcommand{\Tr}{\operatorname{Tr}}

\renewcommand{\(}{\left(}
\renewcommand{\)}{\right)}

\newcommand{\asdef}{=\mathrel{\mathop:}}      

\newcommand{\vphi}{\varphi}

\title{A proper fixed functional for \\ four-dimensional Quantum Einstein Gravity }

\author[a,b]{Maximilian Demmel,}
\author[b]{Frank Saueressig,}
\author[c]{Omar Zanusso}
\affiliation[a]{
PRISMA Cluster of Excellence \& Institute of Physics (THEP), \\
University of Mainz, Staudingerweg 7, D-55099 Mainz, Germany
}
\affiliation[b]{
Institute for Mathematics, Astrophysics and Particle Physics (IMAPP),\\
Radboud University Nijmegen, Heyendaalseweg 135, 6525 AJ Nijmegen, The Netherlands
}
\affiliation[c]{
Theoretisch-Physikalisches Institut, Friedrich-Schiller-Universit\"{a}t Jena, \\
Max-Wien-Platz 1, 07743 Jena, Germany
}
\emailAdd{demmel@thep.physik.uni-mainz.de}
\emailAdd{f.saueressig@science.ru.nl}
\emailAdd{omar.zanusso@uni-jena.de}

\abstract{
Realizing a quantum theory for gravity based on Asymptotic Safety hinges
on the existence of a non-Gaussian fixed point of the theory's
renormalization group flow. In this work, we use the functional
renormalization group equation for the effective average action
to study the fixed point underlying Quantum Einstein Gravity
at the functional level including an infinite number of
scale-dependent coupling constants. We formulate a list of guiding
principles underlying the construction of a partial differential equation
encoding the scale-dependence of $f(R)$-gravity. We show that this
equation admits a unique, globally well-defined fixed functional
describing the non-Gaussian fixed point at
the level of functions of the scalar curvature. This solution is
constructed explicitly via a numerical double-shooting method. In the UV,
this solution is in good agreement with results from polynomial expansions including
a finite number of coupling constants, while it scales proportional to
$R^2$, dressed up with non-analytic terms, in the IR. We demonstrate that its structure is mainly governed by the
conformal sector of the flow equation. The relation of our work to
previous, partial constructions of similar scaling solutions is discussed.
}

\keywords{Quantum Gravity, Asymptotic Safety, Functional Renormalization Group}

\begin{document}
\maketitle
\section{Introduction}
\label{sect.1}
Renormalization group (RG) fixed points are crucial for determining the properties of the underlying quantum field theory.
They are lighthouses on theory space because in their vicinity one can explore
the corresponding quantum field theory via scaling techniques and, eventually,
via conformal field theory methods.
Fixed points also play a crucial role in providing a general notion of renormalizability and ultraviolet completion of a quantum theory.
In fact, in the vicinity of an ultraviolet (UV) fixed point the RG flow can be linearized
and characterized by directions along which it is attracted to or repelled from the fixed point for increasing energies,
thus providing a generalized notion of relevant and irrelevant deformations of the theory.
The UV completeness of the theory is then ensured by finding a given RG trajectory that is attracted towards the fixed point at high energies,
implying that any UV cutoff can be removed in such a way that all essential dimensionless coupling constants remain finite.
Such an UV complete theory is also predictive if the dimension of the critical hypersurface that is attracted towards the fixed point in the UV is finite.
In this case, only a finite number of experiments is required to pinpoint the specific RG trajectory that completes gravity in the deep UV regime.
Depending on whether the fixed point corresponds to a free (Gaussian fixed point) or interacting theory (non-Gaussian fixed point),
the RG trajectories terminating at the fixed point in the UV are termed asymptotically free or asymptotically safe, respectively.
While it is well known that the Gaussian fixed point (GFP) of gravity does not provide a suitable mechanism to renormalize the theory,
the possibility that the theory possesses at least one suitable non-Gaussian fixed point (NGFP) is not speculative and has been explored under various convincing approximations.
In this work we investigate the Asymptotic Safety mechanism for gravity relaxing some of these approximations.
In particular we construct the first consistent and {\it globally well-defined fixed functional} for Quantum Einstein Gravity in four spacetime dimensions.

\subsection{Quantum Gravity via Asymptotic Safety}
\label{sect.1a}

The RG flow of any theory admits a GFP at which the theory itself becomes non-interacting.
In the neighborhood of the Gaussian point it is thus possible to use the methods of perturbation theory to investigate the UV completion of the theory under consideration.
These methods have been applied very successfully to Yang-Mills theories thanks to the property of asymptotic freedom.
However, it has been known for many years that the GFP cannot provide a UV completion mechanism of General Relativity \cite{tHooft:1974bx,Goroff:1985sz,vandeVen:1991gw}.
The question whether gravity can be formulated as a consistent and predictive (asymptotically safe) quantum field theory then becomes closely intertwined with the question whether there are other fixed points of the RG on the gravitational theory space which could provide such a UV completion.
Indeed it has been conjectured already in the late seventies that there exists a suitable NGFP on the gravitational theory space which could render gravity asymptotically safe \cite{Weinberg:1980gg}.

The primary tool for investigating the Asymptotic Safety mechanism has been
a Wilsonian-type functional renormalization group equation (FRGE)
\cite{Wetterich:1992yh,Morris:1993qb,Reuter:1993kw}
\be \label{FRGE}
k \p_k \Gamma_k = \frac{1}{2}  {\rm Tr}  \left[ \left( \Gamma_k^{(2)} + \cR_k \right)^{-1} \, k \p_k \cR_k \right] \,,
\ee
in which $\Gamma_k$ denotes an effective ``average'' action, $\Gamma_k^{(2)}$ is its second variation with respect to the fields, and $\mathcal{R}_k$ is a suitable infrared-regulator
which ensures that the rhs of the equation is finite and peaked around the renormalization group scale $k$.
The effective average action $\Gamma_k$ represents an effective action of the system
in which the UV modes have been integrated out in the path-integral using the scale $k$ as a reference,
thus providing a natural effective action for processes occurring at energies $E \simeq k$.
In the limit $k\to0$ the IR cutoff $\mathcal{R}_k$ vanishes
and the effective average action coincides with the full effective action of the system $\Gamma=\Gamma_{k=0}$.
The flow \eqref{FRGE} is exact in the sense that no approximation is used in its derivation.

Notably, the FRGE allows to compute approximate RG flows of a theory without the need of a small expansion parameter.
For gravity this approach has been pioneered in \cite{Reuter:1996cp} which, by now, has evolved in a very successful research program,
see \cite{Niedermaier:2006wt,robrev,Reuter:2012id,Reuter:2012xf} for reviews.
The key strategy for obtaining non-perturbative information on the RG flow is to project the exact flow
on a subspace spanned by the ``most relevant interactions'' for the physics at hand.
Practically, this strategy is implemented by making a suitable ansatz for $\Gamma_k$
limiting the set of interaction terms to those which are
argued to describe the relevant physical information.
The beta functions, which describe the scale-dependence of the coupling constants contained in the ansatz and drive the approximated RG flow,
are constructed by substituting the ansatz for $\Gamma_k$ into the FRGE and retaining the terms contained in the ansatz itself only.
While these beta functions lack the control parameter $\hbar$ of the loop expansion of perturbation theory \cite{Codello:2013bra},
they are fully non-perturbative in the couplings and thus, for sufficiently big truncations, reliable also away from the GFP.

The projection of $\Gamma_k$ to finite dimensional subspaces has led to a substantial body of evidence supporting the existence of a gravitational NGFP suitable for Asymptotic Safety.
Starting from the Einstein-Hilbert truncation, which contains scale-dependent Newton's and cosmological constants only \cite{Reuter:1996cp,Dou:1997fg,Souma:1999at},
the existence of the NGFP has successively been established on truncations of increasing complexity and field content,
see \cite{Lauscher:2001ya,Reuter:2001ag,Lauscher:2001rz,Lauscher:2002sq} for selected references.
For the case in which the gravitational degrees of freedom are carried by a (Euclidean) spacetime metric, this line of research has already established striking results.
Foremost, all computations corroborated the existence of a NGFP with suitable properties for Asymptotic Safety.
In particular, the NGFP has been put to test and proved robust to the inclusion  of the problematic operators appearing in the perturbative counter-terms in $\Gamma_k$
\cite{Benedetti:2009rx,Benedetti:2009gn}, thus directly showing the feasibility of the program beyond perturbation theory.
Despite this fixed point being associated with an interacting quantum field theory in which the (background) Newton's constant scales with an anomalous dimension $\eta_N = -2$,
classical power-counting may still constitute a good ordering principle for identifying the relevant deformations of the fixed point as shown in \cite{Codello:2007bd,Machado:2007ea,Codello:2008vh,Falls:2013bv}.
In particular the number of relevant parameters of the theory could be as low as three,
implying that, compared to many other quantum field theory models, an asymptotically safe theory of gravity can be very predictive.
Moreover, stringent bounds on the numbers of free matter fields compatible with the Asymptotic Safety mechanism have been obtained in Refs.\ \cite{Dona:2013qba,Dona:2014pla}.
At the level of the formalism, there has also been progress on computing properties and expectation values of the fluctuation fields around a given background \cite{Manrique:2009uh,Manrique:2010mq,Donkin:2012ud,Manrique:2010am,Christiansen:2014raa}.
By now it is appreciated that this type of computations, known as bi-metric computations, may be essential for precision computations of the properties of NGFP such as its critical exponents,
addressing global aspects of gravitational RG flow by, e.g., establishing a suitable $c$-theorem \cite{Becker:2014pea}, or establishing background-independence of the formalism \cite{Becker:2014qya}.

While the case where the gravitational degrees of freedom are carried by metric variables is understood the best, this scenario is not the only way to realize Asymptotic Safety in gravity: a suitable NGFP has been found to occur rather independently of the type of gravitational degrees of freedom. Investigations of the gravitational RG flows based on the ADM-formalism \cite{Manrique:2011jc,Rechenberger:2012dt}, the Einstein-Cartan formalism \cite{Daum:2010qt,Benedetti:2011nd,Daum:2013fu,Harst:2014vca} and the ``tetrad only'' formalism \cite{Harst:2012ni,Gies:2013noa} also identify a NGFP whose properties are strikingly similar to the one encountered in the metric formalism. This is rather remarkable since, while these formulations are equivalent at the classical level, it is far from clear that this equivalence also holds at the quantum level. Moreover, there are also first indications that Asymptotic Safety may be realized in unimodular gravity \cite{Eichhorn:2013xr,Saltas:2014cta,Eichhorn:2015bna}. This leads to the perhaps surprising conclusion that, at the level of finite-dimensional projections, the appearance of a NGFP in gravitational theories is rather commonplace instead of exceptional.\footnote{Notably, Ho\v{r}ava-Lifshitz gravity \cite{Horava:2009uw} also possesses a UV fixed point, which is different from the one underlying the Asymptotic Safety scenario \cite{D'Odorico:2014iha}.
It is conceivable that the Monte Carlo simulations carried out within the Causal Dynamical Triangulations Program \cite{Ambjorn:2012jv,Ambjorn:2012ij,Cooperman:2014sca,Cooperman:2014} are capable of probing both the Ho\v{r}ava-Lifshitz and Asymptotic Safety universality class.
}

\subsection{From fixed points to fixed functions}
\label{sect.1b}
Currently, one of the key conceptual challenges in the Asymptotic Safety program is the leap from approximations for $\Gamma_k$ containing a finite number of scale-dependent couplings to truncations containing an infinite number of coupling constants. In the latter case, $\Gamma_k$ contains scale-dependent functions. Substituting such an ansatz into the exact functional renormalization group equation leads to a (system of) partial differential equations (PDE) which govern the scale-dependence of these functions. The fixed points appearing in finite-dimesional approximations of $\Gamma_k$ are then promoted to fixed functions, namely global $k$-stationary solutions of the PDE.
(Since both fixed points and fixed functions induce fixed functionals, we use the notion of fixed functions in order to stress that we work in the realm of infinitely many scale-dependent coupling constants.)

The simplest setting where these generalizations can be implemented for gravity includes an arbitrary function of the scalar curvature $R$
\be\label{fransatz}
\Gamma_k^{\rm grav}[g] = \int d^dx \sqrt{g} \, f_k(R) \, ,
\ee
where $g$ denotes the (Euclidean) spacetime metric. This so-called $f(R)$-truncation provides
the simplest example for a truncation including a scale-dependent function, since the PDE governing the $k$-dependence of $f_k(R)$ 
can be derived by using a maximally symmetric
background with covariantly constant curvature. 

Starting from the flow equation  derived in \cite{Machado:2007ea}, the existence of fixed functions $f_*(R)$ have been investigated by various groups in $d=3$ \cite{Demmel:2012ub,Demmel:2013myx,Demmel:2014sga} and $d=4$ \cite{Benedetti:2012dx,Dietz:2012ic,Benedetti:2013jk,Dietz:2013sba,Dietz:2015owa} spacetime dimensions. Quite unsettling, the verification of a suitable NGFP at the level of fixed functions turned out to be extremely challenging: while the finite-dimensional computations always produced a suitable UV fixed point regardless of the computational details and setting, up to now the fixed function completing the four-dimensional NGFP in the ansatz \eqref{fransatz} is still elusive. This deficit acted as catalyst to readdress some of the fundamental questions in the functional renormalization group approach to Quantum Einstein Gravity (QEG).

Substituting the ansatz \eqref{fransatz} into the FRGE and projecting on the corresponding subspace leads to a non-linear PDE which is first order in $k$ and third order in the $R$-derivatives. Imposing the condition that the fixed function is $k$-stationary, this equation reduces to a non-linear ordinary differential equation (ODE) of third order. Thus, locally, there is a three-parameter family of solutions to this equation. It was then noticed early-on that the condition for the solution to exist globally may reduce the number of free parameters and may produce isolated fixed functions if there is a balance between the order of the ODE and its singularity structure \cite{Dietz:2012ic}
\be\label{singularitycount}
\mbox{degree of ODE} - \mbox{number of singularities} = 0 \, . 
\ee
Thus any singularity reduces the number of free parameters by one.
While the number of fixed singularities can essentially be read off 
from the ODE, determining the occurrence of movable singularities often requires
studying explicit solutions. Thus, so far, only fixed singularities played a role in restricting the
number of free parameters in the fixed function in the $f(R)$-truncation.
As a important consequence of \eqref{singularitycount} it was then understood that
the initial flow equation contains too many fixed singularities in order to admit global solutions \cite{Dietz:2012ic,Benedetti:2012dx}.
Based both on physical arguments and on the experience maturated during the course of several $f(R)$-type analysis,
we propose the following guiding principles towards the construction of
a suitable fixed function in four-dimensional QEG,
which will turn out to be a set of sufficient conditions for the existence of said fixed function:
\\

\noindent
{\bf 1) The background should admit only one topology.}\\
The initial flow equation has been derived on a maximally symmetric, compact background sphere $S^d$ with positive scalar curvature. A detailed comparison of the flow equations for $f_k(R)$ in the three-dimensional conformally reduced setting established that the topology of background (non-compact with negative scalar curvature or compact with positive scalar curvature) manifestly influences the singularity structure of the PDE. Additionally, the flow equation on a hyperbolic background with negative scalar curvature does not follow from the analytic continuation of the flow equation on a sphere. We thus argue that it is potentially inconsistent to include more than one topological sector in the flow equation avoiding topology changes in the background.
In the sequel, we restrict the construction to the domain $R > 0$ and do not demand that the solutions admit an continuation to the $R \le 0$ region, thereby avoiding that the fix functional has a suitable continuation into a domain with different topology. \\

\noindent
{\bf 2) The path-integral measure should not introduce spurious singularities.}\\
As it turns out, the measure in the gravitational path integral has a crucial effect on the singularity structure of the ODE. Using a linear background field decomposition
\be\label{backgroundsplitt}
g_{\mu\nu} = \gb_{\mu\nu} + h_{\mu\nu}
\ee
together with a standard measure for the metric fluctuations $h_{\mu\nu}$ and evaluating the flow equation based on the transverse-traceless decomposition of $h_{\mu\nu}$ generally introduces fixed singularities from the contributions of the gauge degrees of freedom, ghost terms, and Jacobians arising in the field decomposition.
We believe that these singularities have to be considered as spurious, since their number can be changed at will by varying unphysical parameters.
We will resort to the flow equation derived in \cite{Demmel:2014hla} in order to avoid the spurious poles of the gauge-fixing sector and of the Jacobians of the field decompositions. This framework can be motivated by geometrical considerations, since it corresponds to a trivialization of the gauge-bundle in field space, and is equivalent to a particular simultaneous choice of regulator function $\cR_k$ and of gauge-fixing, which allows for a precise cancellation between the functional traces capturing the gauge-degrees of freedom and ghost contributions.
\\

\noindent
{\bf 3) The asymptotic behavior is captured only by the exact heat-kernel.}\\
The rhs of the FRGE \eqref{FRGE} is a trace involving a (generally complicated) differential operator.
In finite-dimensional truncations, this functional trace is typically evaluated by applying the early-time expansion of the corresponding heat-kernel. This is the natural approach if $\Gamma_k$ is organized via a derivative expansion, for example in powers of the background curvatures. It turns out, however, that at the level of full functional truncations the use of the early-time expansion and the evaluation of the trace based on the exact heat-kernel on $S^d$ leads to a different asymptotic behavior when $k^2 \ll R$. This difference is caused by non-analytic contributions to the heat-kernel, related to the effect that on $S^d$ the diffusing particle may return to its origin multiple times. This effect is not captured by the (analytic) early-time expansion. These terms are, however, essential for reproducing the correct late-time behavior of the heat-kernel (and are yet another manifestation of the compactness of the background manifold). Therefore, we base the evaluation of the operator trace \eqref{FRGE} on the exact heat-kernel including non-analytic contributions to correctly account its asymptotic behavior. \\

\noindent
{\bf 4) Scalar and TT-fluctuations should be integrated out simultaneously.}\\
The initial flow equation has been derived using the Laplacian of the background sphere
for the construction of a covariant regulator.
In the nomenclature of \cite{Codello:2008vh}, this corresponds to a type I cutoff. Our flow equation generalizes this construction by allowing a non-zero endomorphism in the regulator function, using a so-called type II cutoff. Besides making the solutions of the ODE more stable, the freedom of the endomorphism allows to satisfy both the singularity count \eqref{singularitycount} and the correct asymptotic behavior of the solution. Moreover, choosing different endomorphisms in the different sectors of the flow equation allows adjusting the scale at which the lowest eigenmode of the fluctuation field is integrated out. We thus need a physically motivated principle to fix this additional freedom.
We will impose the {\it principle of equal lowest eigenvalues} \cite{Demmel:2014hla}, enforcing that the lowest energy mode in the transverse-traceless and scalar sectors are integrated out at the same value of $k$. Since the order of the highest derivative of $f_k(R)$ differs in the transverse-traceless and scalar sector of ODE, this choice in particular ensures that the order of the ODE does not change on the integration interval.\\

\noindent
{\bf 5) All quantum fluctuations need to integrated out.}\\
A crucial question related to the existence of globally well-defined solutions is the integration range for which the solution ought to exist. The ODE describing the fixed function is typically formulated in terms of the dimensionless curvature $r \equiv \bar{R}/k^2$. For fixed (background) curvature $\bar R$ the interval
 $k \in [0, \infty[$ 
is then mapped to $r \in [\infty, 0]$, so it seems natural to demand that solutions are regular on the entire positive real axis. This is in agreement with the intuition from scalar field theory in a flat background and the requirement that the regulator $\cR_k \propto k^2$ 
vanishes for $k^2 \rightarrow 0$, so that all quantum fluctuations are integrated out. Based on this logic the domain of definition for the fixed function may be inferred by imposing that
\begin{enumerate}
\item[a)] all quantum fluctuations are integrated out,
\item[b)] there are no residual regulator effects in the flow equation.
\end{enumerate}
Depending on the particular choice of regulator, and in particular for Type II cutoffs in combination with the Litim regulator \cite{Litim:2001up}, these requirements may already be satisfied at a finite $r_{\rm max}$. In particular this construction removes the regulator effects not via the ``canonical mechanism'' sending $k \rightarrow 0$ but by the stepfunction containing the eigenvalues of the differential operator becoming zero {once the last fluctuation mode is integrated out}.
In such a case, after the last eigenfluctuation is integrated out, the flow will be driven by the canonical scaling of the function $f(R)$.
Based on this mechanism the $r$-interval, on which the global solution is required to be regular, may be restricted to a finite interval $r \in [0, r_{\rm max}]$,
where $r_{\rm max}$ depends on the choice of endomorphism included in the regulator and on the regulator itself.
If the flow equation is derived using a smooth regulator or the behavior of a smooth regulator is mimicked by applying a smoothing procedure
the contribution of the quantum fluctuations are smeared over the entire half-axis $r \in [0, \infty]$. 
Since we apply the smoothing procedure \cite{Benedetti:2012dx} in the construction of our flow equation, 
we consequently require that a proper fixed function is globally well-defined on this domain.\\

In this work, we establish that these guiding principles lead to a PDE governing the scale-dependence of the ansatz \eqref{fransatz} which gives rise to an {\it isolated, unique and globally well-defined fixed function}, thus reconciling previous results based on finite-dimensional polynomial truncations with a functional ansatz.
While it could have been argued via the counting theorem that only a finite number of solutions was present,
it is in fact amazing to discover that the solution resulting from the application of the above principles is unique.
Let us stress again, however, that our guiding principles constitute a set of sufficient conditions for the existence of said fixed function. At this stage it is unclear whether all principles are strictly necessary for establishing a suitable fixed function. In particular, the list does not incorporate the recent observation \cite{Bridle:2013sra,Dietz:2015owa} that the combination of a flow equation with the equation for the modified split Ward identities may lead to substantial simplifications of the PDEs prescribing the fixed functions of the theory.

The remaining paper is organized as follows. We start by illustrating
the effects of the singularity theorem \eqref{singularitycount} together
with the analytic and numerical techniques used to analyze the existence of fixed functions
 in a simple toy model in Sect.\ \ref{sect.2}. Implementing the guiding principles
discussed above, the ODE governing the existence of fixed functions
in the $f(R)$-truncation is constructed in Sect. \ref{sect.3}.
The properties of the resulting ODE are discussed in Sect.\ \ref{sect.4}
and a numerical study of its solutions is carried out in Sect.\ \ref{sect.5}. 
The central result is the isolated, globally well-defined fixed function shown in Fig.\ \ref{fig:numsol}.
We end with a discussion in Sect.\ \ref{sect.6}.
The analogue of the full fixed function arising
from the conformally reduced approximation
of our flow equation is constructed in Appendix \ref{App:A}.

\section{Isolated global solutions: a toy model}
\label{sect.2}
Before embarking on the quest of constructing our flow equation for $f(R)$-gravity
and its fixed functions, it is useful to illustrate the techniques used
in the analysis by applying them to a simple, completely tractable example.
For this purpose, we analyze the second order ODE
\be\label{boundexample}
y^{\prime\prime}(x) = - \frac{y(x)}{x-1} \, , \qquad y(0) = y_0 \, , \; y^{\prime}(0) = y_1 \, , 
\ee
on the interval $x \in [0, \infty[$. Locally, at $x=0$, the equation admits a two-parameter family of solutions 
which we parameterize by $(y_0, y_1)$.

Inspecting the rhs of eq.\ \eqref{boundexample} one easily sees that the denominator has a root at $x_{\rm sing} = 1$
which constitutes a fixed singularity. A solution with generic initial conditions $(y_0, y_1)$ imposed at $x=0$ will have a diverging second
derivative at $x_{\rm sing}$. A necessary condition for regularity of the rhs is 
\be\label{boundarycond}
\left. y(x)\right|_{x=x_{\rm sing}} = 0 \, , 
\ee
i.e., a globally well-defined $y(x)$ has at least a simple root at $x_{\rm sing}$. Thus it is expected that not all solutions can be extended to global solutions which are well-defined on the entire positive half-axis. The fixed singularity provides a non-trivial boundary condition which puts additional constraints on
the parameters $(y_0, y_1)$ characterizing the local solutions. We now discuss various techniques which allow to analyze this constraint.  \\

\noindent
{\bf The exact analytic solution} \\
The linear ODE \eqref{boundexample} is still simple enough so that its general solution
can be given in terms of modified Bessel functions
\be\label{exactsol}
y(x) = \sqrt{1-x} \, \left( c_1 \, I_1(2 \sqrt{1-x}) + c_2 \, K_1(2 \sqrt{1-x}) \right) \, . 
\ee
\begin{figure}[t]
	\begin{center}
	\includegraphics[width=7cm]{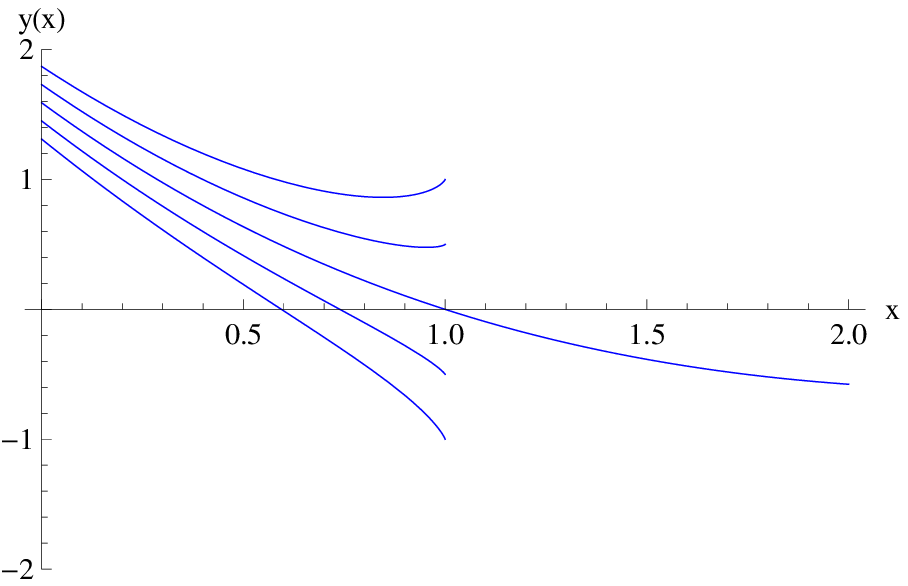} \qquad \includegraphics[width=7cm]{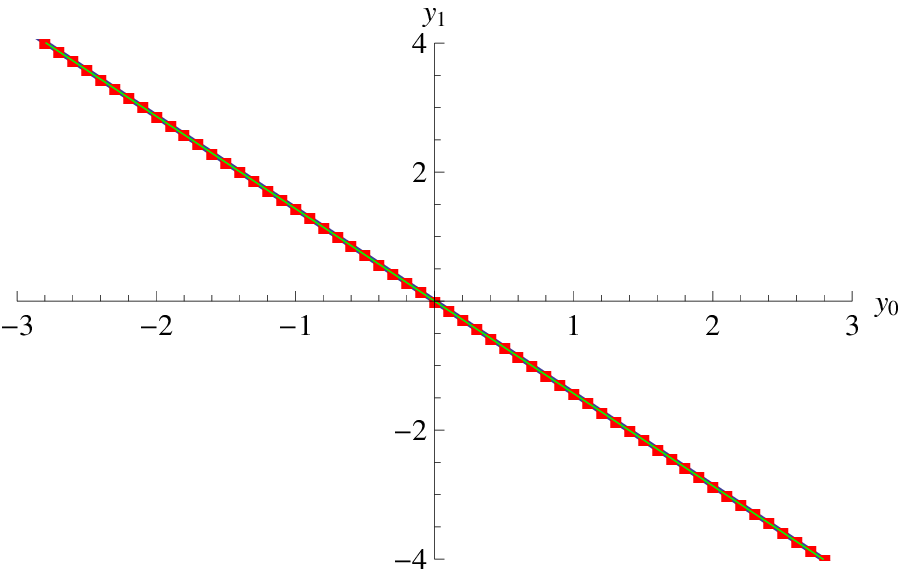}
	\caption{\label{fig.odeexample} Illustration of a one-parameter family of solutions \eqref{exactsol} given by $c_1 = 1$ and various values of $c_2$ (left diagram). The solution $c_2 = 0$ can be extended beyond the fixed singularity $x_{\rm sing} = 1$. For $c_2 \not = 0$ the solutions terminate at the fixed singularity. The right diagram shows the values $(y_0, y_1)$ satisfying the condition \eqref{boundarycond}. The bold blue line is obtained from the exact solution, the red squares indicate the results from the numerical shooting method, and the green line approximates the exact solution
by a polynomial with $N = 12$.}
\end{center}
\end{figure}
Evaluating the solution in the limit $x \rightarrow 1^-$, the term containing $I_1$ vanishes while the $K_1$ term
remains finite. Thus the condition \eqref{boundarycond} implies $c_2 = 0$. Indeed one can explicitly verify that
only solutions with $c_2 = 0$ remain real on $x \in [0, \infty[$: if $c_2 \not = 0$ the $K_1$-term introduces a non-zero imaginary part if $x > 1$.
This property is illustrated in the left diagram of Fig.\ \ref{fig.odeexample}.
The regularity property \eqref{boundarycond} is thus a sufficient condition for the existence of a global real-valued solution in the positive real half-axis.

Expanding \eqref{exactsol} in a Taylor-series at $x=0$ up to linear order gives the relation between the integration constants $c_i$ to the parameters $(y_0, y_1)$
\be\label{dictionary1}
y_0 = c_1 \, I_1(2) + c_2 \, K_1(2) \, , \qquad 
y_1 = - c_1 \left(I_1(2) + I_2(2) \right) - c_2 \left(K_1(2) - K_2(2) \right) \, . 
\ee
Imposing the condition for a regular, global solution, $c_2 = 0$ and eliminating
$c_1$ from the system \eqref{dictionary1} shows that the global solutions
are given by a line in the $(y_0, y_1)$-plane, satisfying
\be\label{globalcon}
y_1 = - y_0 \, \left( 1 + \frac{I_2(2)}{I_1(2)} \right) \approx - 1.433 \, y_0 \, . 
\ee
This line is displayed as the bold blue line in the right panel of Fig.\ \ref{fig.odeexample}.
Thus the interplay between demanding that a solution to exist globally and the fixed singularity of the ODE
reduces the number of free parameters by one. \\

\noindent
{\bf Polynomial expansion at $x=0$} \\
Concerning the ODE describing fixed functions in quantum gravity, it is not likely
that exact, analytic solutions can be found. Thus other techniques to derive the condition
for a global solution \eqref{globalcon} are required. A tractable method approximates
the exact solution by a polynomial of finite order
\be\label{polyapprox}
y(x) \approx \sum_{n=0}^{N} \, y_n \, x^n \,. 
\ee
Substituting this ansatz into \eqref{boundexample} and matching the coefficients multiplying
powers of $x$ up to $x^{N - 2}$ yields $N - 1$ equations relating
the coefficients $y_n$:
\be\label{coeffsys1}
\begin{split}
y_2 = & \, \half \, y_0 \\
y_3 = & \, \tfrac{1}{6} \, \left( y_0 + y_1 \right) \\
\vdots & \\
y_N = & \,  \frac{1}{N(N-1)} \, \sum_{n=0}^{N-2} \, y_n \, . 
\end{split}
\ee 

This system can be used in two ways. Firstly, it can be complemented by the two equations
appearing at orders $x^{N - 1}$ and $x^{N}$:
\be\label{completion1}
\sum_{n=0}^{N-1} \, y_n = 0 \, , \qquad \sum_{n=0}^{N} \, y_n = 0 \, . 
\ee
Combining eqs.\ \eqref{coeffsys1} and \eqref{completion1} then shows that this completion scheme
only leads to the trivial solution $y_n = 0, \forall \, n \in 0, \ldots, N$.

Alternatively, one can follow the strategy of fixing free coefficients by demanding the global existence of
the solution. For this case, it is convenient to cast the system \eqref{coeffsys1} into the form
\be\label{recursion}
y_n = \frac{n-2}{n} \, y_{n-1} + \frac{1}{n(n-1)} \, y_{n-2} \, , \qquad n \ge 2
\ee
from which the $y_n$ can be determined as functions of the parameters $y_0, y_1$ in a recursive 
manner. Fixing $N=12$ and evaluating the regularity condition \eqref{boundarycond} for the polynomial \eqref{polyapprox} then
yields the regularity condition
\be
y_1 = - \frac{1004933203}{701216922}  \, y_0 \, \approx - 1.433 \, y_0 \, . 
\ee
This condition is illustrated by the thin green line in the right panel of Fig.\ \ref{fig.odeexample} and is essentially identical to \eqref{globalcon}. 
Notably, fixing all boundary conditions at $x=0$ (first strategy) or demanding the existence of a globally well-defined solution (second strategy) does not
necessarily lead to the same result.\\

\noindent
{\bf Numerical shooting method} \\
The numerical shooting method approximates the space of initial conditions
spanned by $y_0, y_1$ by a lattice with a suitable lattice spacing. For illustrative
purposes we use $\Delta y_i = 0.1$, corresponding to $4941$ pairs of initial conditions
on the $(y_0, y_1)$-plane shown in Fig.\ \ref{fig.odeexample}. At these lattice points,
the ODE \eqref{boundexample} is then solved numerically on the interval $x \in [0, x_{\rm sing} - \epsilon]$
employing a standard ODE-solver. Practically, we chose $\epsilon = 10^{-3}$.

The numerical solutions provide a map
\be\label{boundarycond2}
\left. (y_0,y_1) \mapsto e(x; y_0, y_1)\right|_{x=x_{\rm sing} - \epsilon} \, . 
\ee
The regularity condition \eqref{boundarycond} corresponds to
 $e(x_{\rm sing} - \epsilon, y_0,y_1) = 0$ and defines a submersion on the space of initial conditions $(y_0,y_1)$. Technically, this condition
is implemented by looking for sign changes in $e(y_0,y_1)$
along the strips of constant $y_0$. The resulting (interpolated) values for $y_1$
indicating a zero of \eqref{boundarycond2} 
are given by the red squares displayed in the right panel of Fig.\ \ref{fig.odeexample}. Fitting a linear
function to the data points yields  
\be
y_1  \approx -1.434 \, y_0 \, .
\ee
This matches the exact result \eqref{globalcon} with three digit precision. Thus all three methods, exact solutions, polynomial approximations and the numerical shooting
method are suitable to quantitatively analyze the constraints implied by the existence of a global solution on the free parameters of the ODE.

\section{The flow equation for $f_k(R)$}
\label{sect.3}
In this section, we construct the PDE encoding the scale-dependence of $f_k(R)$
based on the guiding principles laid down in the introduction. The key results
are eqs.\ \eqref{eq:smoothapprox} and \eqref{eq:fpeq} which constitute the starting points in our search for 
 fixed functions in the subsequent sections.
\subsection{Projecting the RG flow}
\label{sect.31}
The construction of the flow equation \eqref{FRGE} employs the background
field method, splitting the physical metric $g_{\mu\nu}$ into a fixed
background $\gb_{\mu\nu}$ and a fluctuation field $h_{\mu\nu}$
\be
g_{\mu\nu} = \gb_{\mu\nu} + h_{\mu\nu} \, . 
\ee 
When projecting the RG flow onto the ansatz \eqref{fransatz}, it is convenient 
to choose $\gb_{\mu\nu}$ as the metric on the compact four-sphere $S^4$ with
background curvature $\Rb$ and preform a transverse-traceless
decomposition of the metric fluctuations with respect
to this background
 \be
\label{eq:TTdecomposition}
h_{\mu\nu} = h_{\mu\nu}^{\rm T} + \bar{D}_\mu \xi_\nu + \bar{D}_\nu \xi_\mu + 2 \bar{D}_\mu \bar{D}_\nu \sigma  - \frac{1}{2} \gb_{\mu\nu} \bar D^2 \sigma + \tfrac{1}{4} \, \bar{g}_{\mu\nu} \, h \, .
\ee
The component fields consist of a transverse-traceless tensor $h_{\mu\nu}^{\rm T}$ (spin 2), a transverse vector $\xi_\mu$, and the two scalars $\sigma, h$ (spin 0) satisfying the constraints
\eq{
	 \bar{D}^\mu h_{\mu\nu}^{\rm T} =0, \qquad \gb^{\mu\nu} h_{\mu\nu}^{\rm T} =0,\qquad \bar{D}_\mu \xi^\mu=0, \qquad \gb^{\mu\nu} h_{\mu\nu} = h \,  .
}
Following \cite{Demmel:2014hla}, the projected flow for the ansatz \eqref{fransatz} takes the form
\eq{
\label{flowprojected}
\del_t \Gamma_k [\gb]
=
\frac{1}{2} {T}_{(2)}
+
\frac{1}{2} {T}_{(0)}
}
with the explicit expressions for the traces given by
\spliteq{\label{fullequations}
T_{(2)}
&= 6 \, 
\Tr \left[ \left( \left( (3 \alpha + 1)  \bar{R}- 3 P_k \right) f' -  f \right)^{-1} \del_t R_k^{(2)} \right] \, ,
\\
T_{(0)}
&=
16 \, \Tr \left[ 
 \left( \left((3 \beta + 1)\bar{R}-3 P_k\right){}^2 f'' - \left( (3 \beta + 2) \bar{R} - 3 P_k\right) f' +2 f \right)^{-1} \, \del_t R_k^{(0)} \right] \, .
}
Here $t = \ln k/k_0$ is the dimensionless ``renormalization group time'', the primes denote derivatives of $f_k(\Rb)$ with respect to $\Rb$ and
$P_k \equiv \Box + R_k(\Box)$ denotes the regularized kinetic term constructed from the spin-dependent coarse-graining operator $\Box_{(s)} = - \bar{D}^2 + E_{(s)}$.
In the nomenclature of \cite{Codello:2008vh}, the inclusion of non-trivial endomorphisms $E_{(s)}$ in the regulator function gives
rise to a type II regulator. On a spherical background $E_{(s)}$ is proportional to $\Rb$ and we set
\be\label{endomorphisms}
E_{(2)} \equiv \alpha \bar{R}\,, \qquad E_{(0)} \equiv \beta \bar{R}\, . 
\ee
The traces $T_{(2)}$ and $T_{(0)}$ are with respect to the transverse-traceless and scalar fluctuations, respectively.
The particular form of the flow equation \eqref{flowprojected} results from either a particular choice of the path integral 
measure in combination with a special choice of regulator for the gauge degrees of freedom or, alternatively, through
a specific choice of variables on the configuration space of the gravitational path integral \cite{Demmel:2014hla}. 
For vanishing endomorphism the traces coincide with earlier constructions \cite{Codello:2007bd,Machado:2007ea}.

The next step consists in evaluating the traces \eqref{fullequations}. In the case where the flow equation
is expanded in powers of the background curvature $\Rb$ one typically resorts to the local (or early-time) expansion
of the heat-kernel which is truncated at the desired order in $\Rb$. The resulting expansion
is a good approximation in the regime $k^2 \gg \Rb$, but does not give
rise to the correct IR behavior, $k^2 \ll \Rb$, since, by definition, it
does not include the non-analytic terms reflecting the compact nature of the background.
In order to implement construction principle 3 to obtain a flow equation with the correct asymptotic for both $k^2 \gg \Rb$ and  
$k^2 \ll \Rb$, we evaluate the functional traces
as the sum over the eigenvalues of the differential operators.
The spectrum $\lambda_l(d,s)$ and multiplicities $M_l(d,s)$ of the Laplacian on a spherical background 
are well known and summarized in Tab.\ \ref{tab:spectrum}.
\begin{table}[t]
	\centering
	\renewcommand\arraystretch{1.5}
	\begin{tabular}{|c|c|c|l|}
		\hline
		Spin $s$ & Eigenvalue $\lambda_l(d,s)$ & Multiplicity $M_l(d,s)$ & \\ \hline \hline
		0    & $ \frac{l(l+d-1)}{d(d-1)} \bar R $ & $\frac{(2l+d-1)(l+d-2)!}{l!(d-1)!}$  & $l = 0,1,\dots$ \\ \hline
		2    & $\frac{l(l+d-1)-2}{d(d-1)} \bar R$  & $\frac{(d-2) (d+1) (l-1) (d+l) (d+2 l-1) (d+l-3)!}{2 (d-1)! (l+1)!}$ & $l = 2,3,\dots$ \\ \hline
	\end{tabular}
	\caption{Eigenvalues and multiplicities of the Laplacian operator $\Delta \equiv -\bar{D}^2$ acting on fields with spin $s$ on the $d$-sphere \cite{Rubin:1983be,Rubin:1984tc}.}
	\label{tab:spectrum}
\end{table}
Introducing the functions $W_{(i)}$ the functional traces can be written as
\spliteq{\label{traces2}
T_{(2)}
& \equiv \Tr \left[ W_{(2)}(\Delta_{(2)} + \alpha \bar{R}) \right]
& = \sum_{l=2}^{\infty}\; M_l(4,2)\, W_{(2)} \left( \lambda_l(4,2) + \alpha \bar{R} \right)\,,
\\
T_{(0)}
& \equiv \Tr \left[ W_{(0)}(\Delta_{(0)} + \beta \bar{R}) \right]
&=\sum_{l=0}^{\infty}\; M_l(4,0)\, W_{(0)} \left( \lambda_l(4,0) + \beta \bar{R} \right)\,,
}
where $\Delta_{(s)} \equiv - \bar{D}^2$ denotes the standard Laplacian acting on fields with spin $s$.

In order to evaluate the infinite sums explicitly, it is convenient to switch
to dimensionless quantities
\be\label{dimless}
r \equiv \Rb \, k^{-2}\,, \qquad 
f_k(\Rb) \equiv k^4 \varphi_k \left( \Rb k^{-2} \right)\,.
\ee
Moreover, we specify the regulator $R_k$ to be of Litim type \cite{Litim:2001up}
\be
R_k^{(s)}(z) = \left(k^2 - \Box_{(s)} \right) \, \theta_{(s)}(k^2 - \Box_{(s)})
\ee
where $\theta$ is the Heaviside step function. Substituting the traces \eqref{traces2} 
into \eqref{flowprojected} and expressing the result in terms of dimensionless
quantities then yields
\spliteq{\label{eq:spectralsum}
\frac{384 \pi^2}{r^2} \, & \, \left(\dot{\varphi} + 4 \varphi - 2 r \, \varphi' \right) = 
 \sum_{l=2}^\infty\,
M_l(4,2) \, 
\frac{
\tilde{a}_{1,l} \, \varphi' + \tilde{a}_{2,l} \, (\dot{\varphi}' - 2 r \varphi'')
}{
3 \varphi-((3 \alpha+1)  r-3) \, \varphi'
}
\, \theta_{(2)}\\
&\; +
\sum_{l=0}^\infty\,
M_l(4,0) \, 
\frac{
	a_{1,l} \, \varphi' + a_{2,l} \, \varphi'' + a_{3,l} \, \dot{\varphi}' + a_{4,l} \, \left(\dot{\varphi}'' - 2 r \varphi''' \right)
}{
2 \, \varphi
+(3-(3 \beta +2) r) \, \varphi'
+((3 \beta+1)r-3)^2 \, \varphi''
}
\, \theta_{(0)}\,.
}
Here the dots and primes denote derivatives with respect to $t$ and $r$, respectively and we omitted the arguments of $\varphi_k(r)$ to ease readability.
The coefficients in the transverse-traceless sum are
\be\label{eq:lcoeffsTT}
\begin{split}
\tilde{a}_{1,l} = \frac{1}{8} \left( 36 - r (l(l+3)-2 + 12 \alpha  ) \right) \, , \quad 
\tilde{a}_{2,l} = \frac{1}{8} \left( 12 - r (l(l+3)-2 + 12 \alpha -2 ) \right) \, , 
\end{split}
\ee
while the coefficients in the scalar trace are given by
\be\label{eq:lcoeffs}
\begin{split}
a_{1,l} &= \frac{1}{8} \left( 36 - r \, (l(l+3)+12 \beta) \right) \, , \\
a_{2,l} &= \frac{1}{32} \, \left( 432 - 96 \, r \, (3\beta+2) + r^2 \left(l^2 (l+3)^2-144 \beta ^2\right) \right) \, , \\
a_{3,l} &= \frac{1}{8} \left( 12 - r \, (l(l+3)+12 \beta) \right) \, , \\
a_{4,l} &= \frac{1}{32} \, \left( 144 - 96 \, r \, (3\beta+1) + r^2 \left(l^2(l+3)^2 - 8l(l+3) - 48 \beta (3 \beta + 2) \right) \right) \, . \\
\end{split}
\ee
The explicit form of the step functions is given by
\spliteq{
	\label{eq:thetas}
\theta_{(2)} =
\theta\left(1 - \tfrac{1}{12} \, r \, \left(l(l+3)-2+ 12 \alpha \right) \right)\,,
\quad
\theta_{(0)} =
\theta\left(1- \tfrac{1}{12} \, r \, \left(l(l+3) + 12\beta \right)\right)\,.	
}

Eq.\ \eqref{eq:spectralsum} provides a very explicit illustration how the FRGE \eqref{FRGE} works:
the quantum fluctuations are organized with respect to the eigenfunctions of the background Laplacian.
Each eigenmode thereby gives a specific contribution to the trace 
whose weight is determined
through the Hessian $\Gamma_k^{(2)}$. From the step functions \eqref{eq:thetas} one explicitly sees
that lowering $k$ (which, for fixed background curvature $\Rb$, corresponds to increasing $r$) the eigenmodes
are integrated out consecutively starting from fluctuations with large values $l$: once $k^2$ crosses
the corresponding eigenvalues the step-functions ensure that the contribution of the associated spherical harmonic
drops out from the rhs of the flow equation and does no longer contribute to the running of $\varphi_k$.

The structure of the step functions \eqref{eq:thetas} also clarifies the role of the endomorphisms $E_{(s)}$.
The parameters $\alpha$ and $\beta$ allow to adjust the value $k_{(s),{\rm term}}$ at which the last fluctuation
mode of the corresponding trace is integrated out. Substituting the $l$-value of the lowest eigenmode
and equating the argument of $\theta_{(s)}$ to zero yields
\be\label{rterm}
r_{(2), {\rm term}} = \frac{3}{2+3\alpha} \, , \qquad r_{(0), {\rm term}} = \frac{1}{\beta} \, .
\ee 
For $r > r_{(s), {\rm term}}$ the corresponding trace vanishes identically. The 
condition that both the spin-two and scalar trace are completely integrated
out at the same value $r_{\rm term}$ can be implemented by requiring that the
coarse-graining operators $\Box_{(s)}$ have equal lowest eigenvalues (ELE). Equating $r_{(2), {\rm term}}$ and
$r_{(0), {\rm term}}$ shows that the ELE condition is satisfied along the line
\eq{\label{ELE}
\alpha = \beta - \frac{2}{3}\, . 
}
In the following, we will implement this condition thereby realizing the guiding principle 4. 
This choice is motivated on physical grounds, since it implies that
all fluctuations are integrated out at one fixed value of $k$. It turns out
that the ELE-condition can consistently be imposed with all other guiding principles. 

In principle the endomorphisms can also be used to integrate out all fluctuation modes on a compact
interval by selecting $r_{(2), {\rm term}} = r_{(0), {\rm term}} < \infty$. For $r > r_{(0), {\rm term}}$
the rhs of the flow, encoding the quantum effects, becomes trivial and the flow equation reduces to the
classical scaling relation. This was the strategy implemented in \cite{Demmel:2014sga},
which constructed fixed functions on the compact interval $0 \le r \le r_{(0), {\rm term}} = 6$. While
this compactification simplifies the numerical analysis, it will not be implemented here and
we will consider fixed functions which are well-defined on the entire positive half-axis $0 \le r < \infty$.

\subsection{Smoothing the staircase}
A feature that makes \eqref{eq:spectralsum} particularly difficult to work with is that the PDE possesses an infinite number of discontinuities. At the points
\spliteq{
	r_{2,l}  = \frac{12}{12 \alpha +l (l+3)-2},\; l\geq 2,
	\qquad
	r_{0,l} &= \frac{12}{12 \beta +l (l+3)},\; l\geq 0,
}
where $k^2$ crosses an eigenvalue of $\Box_{(s)}$, 
the step functions \eqref{eq:thetas} remove the corresponding eigenmode and the
rhs of \eqref{eq:spectralsum} changes discontinuously. This feature is particular to the use of the Litim regulator and would be absent in the case of a smooth (e.g.\ exponential) regulator. In order to arrive at a smooth PDE we follow the strategy of \cite{Benedetti:2012dx} and approximate this staircase-behavior by a smooth function. This is achieved as 
follows. Firstly, we observe that, for $r > 0$, the sum over eigenvalues is truncated to a finite sum. 
In order for a term to contribute to this series, the corresponding argument of the
theta-function has to be positive. This gives an $r$-dependent upper boundary $N_r$ on the number of terms in the finite sum
\eq{
	12-r (12 \alpha +l (l+3)-2) > 0 \iff l< \frac{\sqrt{r ((17-48 \alpha ) r+48)}-3 r}{2 r} \asdef N_{(2)}(r)
}
for the tensor modes and
\eq{
12 - l (3 + l) r - 12 r \beta >0 \iff l < \frac{\sqrt{3} \sqrt{-16 \beta  r^2+3 r^2+16 r}-3 r}{2 r} \asdef N_{(0)}(r)
}
for the scalar modes.
Again, it can be seen that only for $r=0$ infinitely many terms are contributing.
Regarding the $l$-dependence, the spectral sum contains only a polynomial in $l$, as can be seen in eq.\ \eqref{eq:lcoeffs}.
For $r>0$ each monomial can be summed using the relation
\eq{\label{stepapprox}
	\sum_{n=1}^{M} n^k = \sum_{j=0}^{k} \binom{k}{j} \frac{B_{k-j}}{j+1} M^{j+1}\, ,\qquad B_1 = -\frac{1}{2}\,,
}
with Bernoulli numbers $B_n$.
Substituting $M\mapsto N_{(i)}(r) $ yields a smooth approximation bounding the staircase-function from above.
A smooth approximation that bounds the sum from below is obtained by a sum from $n=1$ to $n=N_r - 1$.
In practice, we approximate the staircase by the average of this two approximations, setting
\eq{\label{average}
	\sum_{n=1}^{N_r} n^k \mapsto \frac{1}{2}\( \sum_{n=1}^{N_r} n^k  + \sum_{n=1}^{N_r-1} n^k \)\,.
}

Applying the summation formula \eqref{average} to the rhs of the flow equation \eqref{eq:spectralsum} yields the desired
smooth PDE govering the scale-dependence of $\varphi_k(r)$
\spliteq{\label{eq:smoothapprox}
\dot{ \varphi} + 4 \varphi - 2 r \varphi'
&=
\frac{
\tilde{c}_1 \varphi' + \tilde{c}_2 (\dot{\varphi}' - 2 r \varphi'')
}{
3 \varphi-((3 \alpha +1)r-3) \varphi'}
\\
&\qquad+
\frac{
c_1 \varphi' + c_2 \varphi'' + c_3 \dot{\varphi}' + c_4\left(\dot{\varphi}'' - 2 r \varphi''' \right)
}{
2\varphi
+(3-(3 \beta +2) r) \varphi'
+(3 \beta  r+r-3)^2 \varphi''
} \, . 
}
The coefficients $c_i$ and $\tilde c_i$ are polynomials in $r$ and depend on the endomorphisms
\spliteq{
\label{ctdef}
\tilde{c}_1
&=
-\tfrac{5 (6 \alpha  r+r-6) \left(\left(18 \alpha ^2+9 \alpha -2\right) r^2-18 (8 \alpha +1) r+126\right)}{6912 \pi ^2}
\\
\tilde{c}_2
&=
-\tfrac{5 (6 \alpha  r+r-6) ((3 \alpha +2) r-3) ((6 \alpha -1) r-6)}{6912 \pi ^2}
}
and
\spliteq{
\label{cdef}
c_1
&=
-\tfrac{((6 \beta -1) r-6) \left(\beta  (6 \beta -1) r^2+(10-48 \beta ) r+42\right)}{2304 \pi ^2}
\\
c_2
&=
-\tfrac{((6 \beta -1) r-6) \left(\beta  \left(54 \beta ^2-3 \beta -1\right) r^3+\left(270 \beta ^2+42 \beta -35\right)
   r^2-39 (18 \beta +1) r+378\right)}{4608 \pi ^2}
\\
c_3
&=
-\tfrac{(\beta  r-1) ((6 \beta -1) r-6)^2}{2304 \pi ^2}
\\
c_4
&=
\tfrac{(\beta  r-1) ((6 \beta -1) r-6)^2 ((9 \beta +5) r-9)}{4608 \pi ^2} \, . 
}

The RG flow of the function $\varphi_k(r)$ is governed by the PDE \eqref{eq:smoothapprox}.
Within this type of functional truncations, fixed points are promoted to fixed functions $\varphi_*(r)$
given by globally well-defined, $k$-stationary solutions of the PDE.
Dropping all $t$-derivatives from \eqref{eq:smoothapprox} then yields a third-order ODE for the fixed function $\varphi_*(r)$
\spliteq{\label{eq:fpeq}
	4 \varphi - 2 r \varphi'
	&=
	\frac{
		\tilde{c}_1 \varphi' -2 \tilde{c}_2 r \varphi''
	}{
	3 \varphi-(3 \alpha  r+r-3) \varphi'
}
+
\frac{
	c_1 \varphi' + c_2 \varphi''  -2 c_4  r \varphi'''
}{
(3 \beta  r+r-3)^2 \varphi''+(3-(3 \beta +2) r) \varphi'+2 \varphi
}
}
with coefficients given in \eqref{ctdef} and \eqref{cdef}.
Since there is no risk of confusion we will omit the asterisk and denote the fixed function only by $\varphi(r)$.

One particular effect of the smoothing procedure applied in this subsection is that rhs of the flow equation is non-zero on the entire interval $r \in [0, \infty[$ even if the endormorphisms are chosen in such a way that the quantum fluctuations in the initial flow equation \eqref{eq:spectralsum} are integrated out on a finite $r$-interval. Thus, implementing the guiding principle 5, we insist that a ``globally well-defined'' solution is required to be regular on the full interval $r \in [0, \infty[$. 

\subsection{Conformally reduced flow equation}
Based on the full flow equation \eqref{eq:smoothapprox}, it is straightforward to write down the ``conformally reduced'' approximation where the RG flow is solely driven by fluctuations of the scalar mode. Eliminating the contributions from the transverse-traceless fluctuations by setting $\tilde c_1$ and $\tilde c_2$ to zero yields
\spliteq{\label{eq:smoothapproxCR}
	\dot{ \varphi} + 4 \varphi - 2 r \varphi'
	&=
\frac{
	c_1 \varphi' + c_2 \varphi'' + c_3 \dot{\varphi}' + c_4\left(\dot{\varphi}'' - 2 r \varphi''' \right)
}{
2\varphi_k
+(3-(3 \beta +2) r) \varphi'
+(3 \beta  r+r-3)^2 \varphi''
}\,,
}
with coefficients \eqref{cdef}.
Setting all $t$-derivatives equal to zero yields the fixed point equation
\spliteq{\label{eq:fpeqCR}
	4 \varphi - 2 r \varphi'
	&=
\frac{
	c_1 \varphi' + c_2 \varphi''  -2 c_4  r \varphi'''
}{
(3 \beta  r+r-3)^2 \varphi''+(3-(3 \beta +2) r) \varphi'+2 \varphi
}
}
which is the analogue of the ODE determining the fixed functions in the full system \eqref{eq:fpeq}. Comparing the ODE's
of the conformally reduced and full system one finds that they share all the characteristic features: both settings give rise to a non-linear ODE of third order. Moreover,
the coefficient multiplying the highest derivative of $\varphi$ is identical, indicating that both equations possess the same fixed singularities. 
On this basis, one may expect that both equations also share the same number of globally well-defined solutions.
This expectation will be confirmed in Appendix \ref{App:A} where eq.\ \eqref{eq:fpeqCR} will be analyzed in detail.

\section{Analytical properties of the stationary flow equation}
\label{sect.4}
The ODE \eqref{eq:smoothapprox} encoding the fixed functions of QEG in the
$f(R)$-approximation is a complicated non-linear third order differential equation
making the construction of exact analytic solutions a difficult task. Nevertheless
some features of the ODE can be accessed by analytic methods. We start with analyzing the fixed singularities in Sect.\ \ref{sect:4.1} before
zooming into the asymptotic IR behavior of solutions in Sect.\ \ref{sect:4.2}.

\subsection{Singularity structure}
\label{sect:4.1}
Being of third order, the ODE \eqref{eq:smoothapprox} gives rise to a three-parameter family of solutions which exist at least locally. 
The condition that such a local solution can be extended to a global solution, existing for all $r\geq 0$, gives rise to constraints on these parameters.
If the ODE contains fixed singularities, global solutions have to fulfill regularity conditions at these points
in order to be able to pass through the singularity. This constrains the space of solutions and generically one expects that each fixed singularity reduces the number of free parameters by one.
The existence of isolated, globally defined solutions generally requires
a total number of three singularities. This condition is met, for
example, if the ODE has three fixed singularities in its domain of
validity.

\begin{figure}[t]
	\begin{center}
		\includegraphics[width=0.7\textwidth]{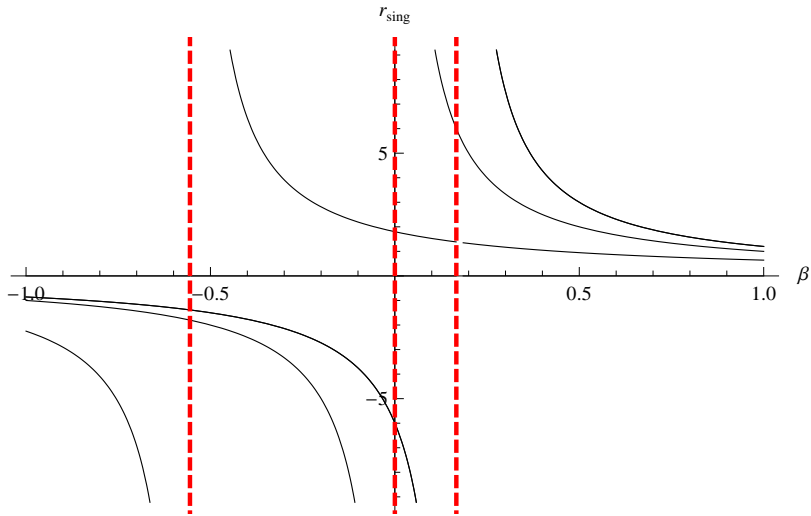}
		\caption{\label{fig:rvsbeta}
			Position of the roots $r_{i, {\rm sing}}$ of the coefficient $c_4$, eq.\ \eqref{cdef}, as a function of the endomorphism $\beta$. At the dashed vertical lines
			one of the roots is located at infinity, implying that the number of fixed singularities on the positive half-axis changes at these values of $\beta$.
		}
	\end{center}
\end{figure}

In order to determine its fixed singularities, the ODE \eqref{eq:smoothapprox} is cast into normal form by solving for $\varphi'''(r)$.
One then sees that the 
coefficient $c_4$ multiplying $ r \varphi'''(r)$ 
determines the fixed singularities of the ODE:
zeros of $c_4$ constitute points where the ODE develops a pole.
Due to the factor $r$ multiplying $\varphi'''(r)$, one pole is always real and located at $r=0$.
This pole originates from switching to dimensionless quantities and the fact that $R$ is not dimensionless.
The coefficient $c_4$ is a forth order polynomial in $r$, giving rise to four additional roots.
Thus set of fixed singularities is given by
\eq{\label{eq:sing}
r_{{\rm sing},1} = 0 \, , 
	&&
	r_{{\rm sing},2} = \frac{9}{5+9\beta} \, , 
	&&
	r_{{\rm sing},3} = \frac{1}{\beta} \, , 
	&& r_{{\rm sing},4,5} = \frac{6}{6\beta-1} \, . 
}
The position of the roots $r_{{\rm sing},i}$, $i = 2,3,4,5$ depends on the endomorphism $\beta$ and is shown in Fig.\ \ref{fig:rvsbeta}.
Notably, $r_{\mathrm{sing},2}$ diverges at $\beta = -\tfrac{5}{9}$ and the double pole $r_{\mathrm{sing},4,5}$ is shifted to infinity if $\beta = \tfrac{1}{6}$,
so that these are distinguished choices for the endomorphism.
The analysis in Fig.\ \ref{fig:rvsbeta} shows that 
that for $\beta = 0$ the ODE has two fixed singularities at finite, positive values $r$.
By choosing the endomorphism in the interval $0<\beta \leq \tfrac{1}{6}$ the number of fixed singularities is increased to three.
For $\beta > \tfrac{1}{6}$ the number of fixed singularities further increases to four.
Thus the inclusion of the endomorphism allows to control the singularity structure
of the ODE to a certain extend.

Based on the singularity counting argument 
presented in Sect.\ \ref{sect.1b},
we then select $\beta$ such that the number of fixed
singularities matches the order of the ODE.
In anticipation of the numerical analysis, we will impose  the condition of ELE and choose
\be\label{eq:endochoice}
\beta = \frac{1}{6} \, , \qquad \alpha = - \frac{1}{2} \, . 
\ee
In this case, the coefficient $c_4$ reduces to a third order polynomial which considerably simplifies the numerical analysis of
the ODE computations.

\subsection{Asymptotic IR behavior of the solutions}
\label{sect:4.2}
The second property which can be accessed analytically is the scaling 
of the solutions $\varphi(r)$ in the IR limit $r\to \infty$.
In order to zoom into this asymptotic region, we rescale $r$ and $\varphi$
by a constant scale parameter $\epsilon$,
\spliteq{
	\label{eq:rescalings}
r = \epsilon^{-a} \tilde{r}\,, 
\qquad
 \varphi(r) =  \epsilon^{-b} \tilde{\varphi}(\tilde{r})\,,
 \qquad
 a > 0\,,
}
and require that the new variables $\tilde{r}$ and $\tilde{\varphi}$ are strictly of the order one.
Since $a >0$ the limit $\epsilon\to 0$ corresponds to $r  \to \infty$. Retaining the
leading terms in $\epsilon$ only, eq.\ \eqref{eq:fpeq} simplifies
to
\eq{
\label{eq:fpeqrescaled}
\frac{1}{\epsilon^b}\;
\left(
4 \tilde{\varphi} -2\tilde{r}\tilde{\varphi}'
\right)
=
\frac{1}{\epsilon^{2a}}\; \frac{5 \tilde{r}^3}{864 \pi ^2} \, 
\frac{ 2 \, \tilde{r} \, \tilde{\varphi }''-\tilde{\varphi }'}{ \tilde{r} \,
	\tilde{\varphi }'+6 \, \tilde{\varphi }}
\, ,
}
where $\tilde \varphi = \tilde \varphi(\tilde r)$ and the primes denote derivatives with respect to the argument.
The lhs is independent of the scale parameter $a$, which can be traced back to the fact that the lhs of eq.\ \eqref{eq:fpeq}
is invariant under rescalings of the form $r\mapsto \lambda r, \varphi \mapsto \varphi$.
Similarly, the rhs is independent of the scale parameter $b$ reflecting that the rhs of eq.\ \eqref{eq:fpeq},  is invariant with respect
to the rescaling $\varphi\mapsto \lambda \varphi, r \mapsto r$. Notably, the rhs of \eqref{eq:fpeqrescaled} receives contributions
from the transverse-traceless sector only. The choice $\beta = 1/6$ reduces the order of polynomials $c_i$ so
that the contribution of the scalar sector is sub-leading in the large-$r$ limit.

Depending on the ratio of $a$ and $b$, eq.\ \eqref{eq:fpeqrescaled} exhibits
three different scaling behaviors: classical scaling for $b > 2a$,
quantum scaling for $b < 2a$ and balanced scaling $b=2a$. These
cases will be discussed below.

\medskip

\noindent
{\bf a) classical scaling $b>2a$} \\
For $b>2a$, the scaling behavior implied by eq.\ \eqref{eq:fpeqrescaled} is 
dominated by its classical lhs 
\eq{
	4 \tilde{\varphi} -2\tilde{r}\tilde{\varphi}' =0\,.
}
This equation is solved by
\eq{\label{asymclass}
	\tilde{\vphi}(\tilde{r}) = A \tilde{r}^2\,,
	}
with $A$ being a free integration constant. \\

\noindent
{\bf b) quantum scaling $b<2a$} \\
In this case the asymptotic behavior entailed in \eqref{eq:fpeqrescaled} is dominated by the rhs encoding 
the quantum fluctuations. Disregarding potential singularities caused by the denominator, the
scaling is governed by the linear equation
\be
2 \tilde{r} \tilde{\varphi }''-\tilde{\varphi }' = 0 \, ,
\ee
which entails that
\eq{
\tilde{\varphi}(\tilde{r})
=
A_1 \tilde{r}^{\frac{3}{2}} + A_2\,,	
}
with $A_1$ and $A_2$ free integration constants. Notably, the existence of the $\tilde{\varphi}(\tilde{r}) = A_1 \tilde{r}^{\frac{3}{2}}$
solution is not intrinsic to the special choice of endomorphisms used to derive \eqref{eq:fpeqrescaled} but appears for generic values $\alpha$ and $\beta$ where the scalar sector also contributes to the asymptotic IR scaling. \\

\noindent
{\bf c)  balanced scaling $b=2a$} \\
In this case, the classical and quantum terms balance and eq.\ \eqref{eq:fpeqrescaled}
entails that
\eq{
	\label{eq:balanced}
	4 \tilde{\varphi} -2\tilde{r}\tilde{\varphi}'
=
 \frac{5 \tilde{r}^3}{864 \pi ^2} \, 
\frac{ 2 \, \tilde{r} \, \tilde{\varphi }''-\tilde{\varphi }'}{ \tilde{r} \,
	\tilde{\varphi }'+6 \, \tilde{\varphi }} \, . 
}
Solving this non-linear equation exactly is beyond the scope of this work. Instead we check if it admits a scaling solution of the form $\tilde{\vphi}(\tilde{r})= A \tilde{r}^a$.
Inserting this scaling ansatz into eq.\ \eqref{eq:balanced} yields
\eq{
2(2-a)A \tilde{r}^a
=
\frac{5 a (2 a-3)}{864 \pi ^2 (a+6)}\tilde{r}^2	\,.
}
For this equation to be satisfied, $\tilde{r}$ has to have the same power on both sides implying classical scaling $a=2$.
For this choice the lhs vanishes, however, so that $a=2$ does not constitute to a consistent solution.

The structure of \eqref{eq:balanced} suggests modifying the scaling ansatz including a logarithmic
term
\eq{
	\tilde{\vphi}(\tilde{r})
	=
	A \tilde{r}^2 \left(1 - \eta  \log (\tilde{r})\right)\,.
}
Inserting this modified ansatz into eq.\ \eqref{eq:balanced} leads 
to a consistent scaling equation. Expanding the result to linear order in $\eta$ 
then fixes the coefficient in terms of $A$
\be
\eta = \frac{40 }{55296 \pi ^2 A+95} \, . 
\ee
Also, in this case the appearance of a logarithmic contribution in the scaling solution
constitutes a generic feature and is not related to the particular choice of endomorphisms.

The scaling solutions obtained in the cases a), b), and c) constitute the three admissible asymptotic behaviors of potential solutions of the full ODE in the IR. A priori it is unclear which phase is realized by a potential fixed function. Nevertheless, this scaling analysis provides important information on the IR dynamics of potential fixed functions in QEG.

\section{Numerical construction of the fixed function}
\label{sect.5}
In this section, we use a numerical shooting method to construct solutions of the ODE \eqref{eq:fpeq} which are regular at the fixed singularities. The central result of this construction is the unique, isolated and globally well-defined fixed function shown in Fig.\ \ref{fig:numsol}.
\subsection{Implementing regularity at the fixed singularities}
\label{sect:5.1}
Our starting point is the ODE \eqref{eq:fpeq} for the specific choice of endomorphisms \eqref{eq:endochoice}.
The analysis \eqref{eq:sing} shows that in this case, the ODE possesses three fixed singularities located at finite values $r$:
\eq{\label{eq:poleposition}
r_{1, {\rm sing}} = 0\, ,&& r_{2, {\rm sing}} = \frac{18}{13}\,, && r_{3, {\rm sing}} = 6\,.
}
Regularity of the solution at these points imposes non-trivial boundary conditions on $\varphi(r_i)$ and its derivatives.
Casting the ODE into normal form by solving for $\varphi'''$, and expanding the resulting rhs
at one of the fixed singularities yields a Laurent expansion of the form
\eq{\label{eq:laurentexp}
\varphi'''(r) = 
\frac{
e_i(\varphi''(r_i),\varphi'(r_i),\varphi(r_i),r_i)
}{
r-r_i
}
+ \textrm{regular terms}\, ,
}
with the residues $e_i$ given by
\spliteq{\label{eq:BCs}
e_1
&=
\frac{
7}{\varphi '+\varphi } \left(\tfrac{13}{27} \varphi \, \varphi ' + \tfrac{1}{2} \varphi \, \varphi '' + \tfrac{2}{3} (\varphi ')^2 + \tfrac{13}{6}\, \varphi '  \varphi ''  \right)
-
\tfrac{256\pi^2}{9}  \varphi   \left(9 \varphi ''+3 \varphi '+2 \varphi \right)
\,,
\\[1.2ex]
e_2
&= \tfrac{1}{16 \varphi '+13 \varphi}
\left( \tfrac{46328}{2535} \varphi \, \varphi '' - \tfrac{400489}{3510} \, \varphi \varphi' +  \tfrac{14837}{2535} (\varphi ')^2 - \tfrac{2468992}{32955}\, \varphi '  \varphi '' + \tfrac{456000}{28561} \left(\varphi ''\right)^2
\right)
\\
&\qquad+
\tfrac{256 \pi ^2}{7605} \left(13 \varphi -9 \varphi '\right)  \left(72 \varphi ''-39 \varphi '+169 \varphi \right)
\,,
\\[1.2ex]
e_3
&=
\tfrac{1}{2 \varphi '+\varphi}  
\left( \tfrac{73}{10}\, \varphi  \varphi' - \tfrac{18}{5}\, \varphi  \varphi'' - \tfrac{207}{5} ( \varphi ')^2  + \tfrac{594}{5}\, \varphi '  \varphi'' \right) 
-\tfrac{
512 \pi ^2}{15} \left(\varphi -3 \varphi '\right)
\left(18 \varphi ''-6 \varphi '+\varphi \right)
\,.
}
Here it is implicitly understood that $\varphi(r)$ and its derivatives appearing in $e_i$ are evaluated at $r_{i, {\rm sing}}$.
Keeping $\varphi'''(r)|_{r = r_i}$ finite then requires $e_i = 0$. A globally well-defined solution
has to satisfy the resulting boundary condition (BC) at all three singular points \eqref{eq:poleposition}.
On this basis, we expect that the BCs fix all three free parameters characterizing the solutions of the ODE locally, so that the set of globally well-defined solutions is discrete.

The next step consists in characterizing the solutions of the ODE by free parameters. Imposing that $\varphi(r)$
satisfies the regularity conditions \eqref{eq:BCs} ensures that the corresponding solution
can be expanded in a Taylor series at $r_{i, {\rm sing}}$ since all its derivatives remain regular.
In particular a fixed function has an analytic expansion in the UV ($r = 0$), so that the solution
locally takes the form
\be\label{eq:gexp}
\varphi_1(r) = g_0 + g_1 r + \sum_{n=2} g_n \, r^n\, . 
\ee
Substituting this ansatz into the ODE and imposing regularity at $r_{1, {\rm sing}}$ by
requiring $e_1 = 0$ uniquely fixes the constants $g_n$, $n \ge 2$ in terms of $(g_0, g_1)$.
In particular,
\be
g_2 = \frac{7 g_1 \left(13 g_0 + 18 g_1 \right) + 768 \pi^2 g_0 \left(2 g_0^2 - 5 g_0 g_1 - 3 g_1^2 \right)}{13824 \pi^2 g_0 (g_0 + g_1)- 63 (3 g_0 + 13 g_1)} \, . 
\ee
Thus, implementing the regularity condition $e_1$, we obtain a two-parameter family of (local) solutions specified by $(g_0, g_1)$.

In a similar fashion, one can perform an analytic expansion of $\varphi(r)$ at the other singular points
\bea
\label{eq:expansions}
\varphi_2(r) 
&=& b_0 + b_1 (r-r_2) + \sum_{n=2} b_n (r-r_2)^n\, ,
\\ \label{eq:cexp}
\varphi_3(r) 
&=& c_0 + c_1 (r-r_3) + \sum_{n=2} c_n (r-r_3)^n\, .
\eea
Combining the ODE with the corresponding regularity condition again fixes the the coefficients $b_n$ and $c_n$ with $n \ge 2$
in terms of $(b_0, b_1)$ and $(c_0, c_1)$, respectively. For the $c_n$ this solution is again unique, while the occurrence of $(\varphi'')^2$
in $e_2$ makes the solution $b_2(b_0, b_1)$ double-valued. In the following we will use the parameterization $(g_0, g_1)$
to characterize potential global solutions, which then already satisfy regularity at $r_{1, {\rm sing}} = 0$.

\subsection{Extending the local solutions away from $r=0$}
\label{doubleshooting}
The next step consists in extending the two-parameter family of solutions \eqref{eq:gexp} to the interval $r \in [0, \infty[$
implementing the boundary conditions \eqref{eq:BCs}. In contrast to the example discussed in Sect.\ \ref{sect.2},
we do not have an analytic solution of the ODE at our disposal. Moreover, the polynomial expansion is difficult to 
implement to sufficiently high order due to the complexity of the ODE. Therefore, we rely on the numerical shooting
method illustrated at the end of Sect.\ \ref{sect.2}. Since we need to implement the boundary conditions
at $r_{2, {\rm sing}}$ and $r_{3, {\rm sing}}$, we divide the positive half-axis into three intervals 
\begin{align}
\label{intervals}
I_1 \equiv ]0,r_2[\,, \quad I_2 \equiv ]r_2,r_3[ \quad \textrm{and } I_3 \equiv ]r_3,\infty[ \,.
\end{align}
The shooting algorithm is applied twice: firstly, the solution is constructed on the interval $I_1$. Solutions meeting the boundary condition $e_2 = 0$ are subsequently extended into the interval $I_2$ where the condition $e_3 = 0$ is tested. Finally, solutions satisfying all three boundary conditions are numerically extended onto $I_3$. 
Owing to the non-linearity of the ODE, a solution might also terminate in a movable singularity so
that it does not extend up to the fixed singularities marking the upper boundaries of the integration intervals. 
 
Practically, we implement this search strategy for fixed functions as follows.
Because of the fixed singularity at $r_{1, {\rm sing}} =0$, the initial conditions
for the numerical integration need to be imposed at $r = \epsilon_1$ rather
than at $r=0$. The initial values for $\vp(r)|_{r = \epsilon_1}$ and its derivatives
in terms of the free parameters $(g_0, g_1)$ are obtained from
the expansion \eqref{eq:gexp}, yielding
\spliteq{\label{eq:inits1}
\vp(\epsilon_1)	&=  \varphi_1(\epsilon_1; g_0,g_1)\,,
\quad
\vp'(\epsilon_1)	=  \varphi_1'(\epsilon_1; g_0,g_1)\,,
\quad
\vp''(\epsilon_1)	=  \varphi_1''(\epsilon_1; g_0,g_1)\,.
}
We then discretize the $(g_0,g_1)$-plane by a lattice and use the initial conditions obtained
at each lattice point to numerically integrate the ODE
from $\epsilon_1$ to $r_{2, {\rm sing}}-\epsilon_2$. This defines a two-parameter family
of solutions
\be\label{numsol1}
\varphi(r; g_0, g_1) \, , \qquad r \in [\epsilon_1, r_{2, {\rm sing}} - \epsilon_2] \, . 
\ee
Evaluating the BC $e_2$, eq.\ \eqref{eq:BCs}, based on these solutions provides a map
\eq{
(g_0,g_1) \mapsto e_2(r_2-\epsilon_2;g_0,g_1) \, .
}
\begin{figure}[t]
\begin{center}
\includegraphics[width=0.5\textwidth]{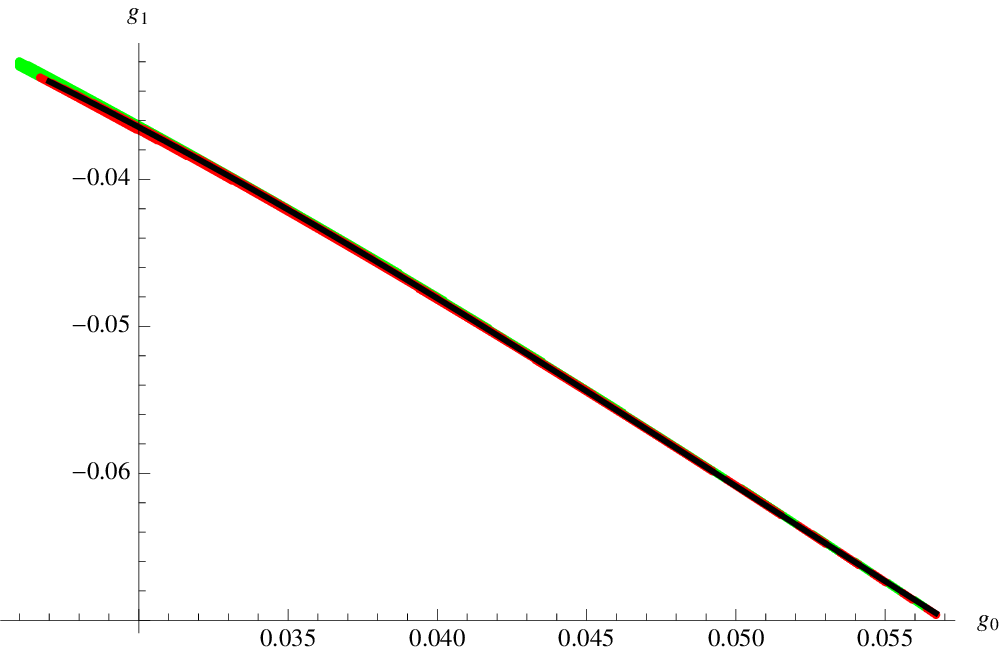} \\
\includegraphics[width=0.48\textwidth]{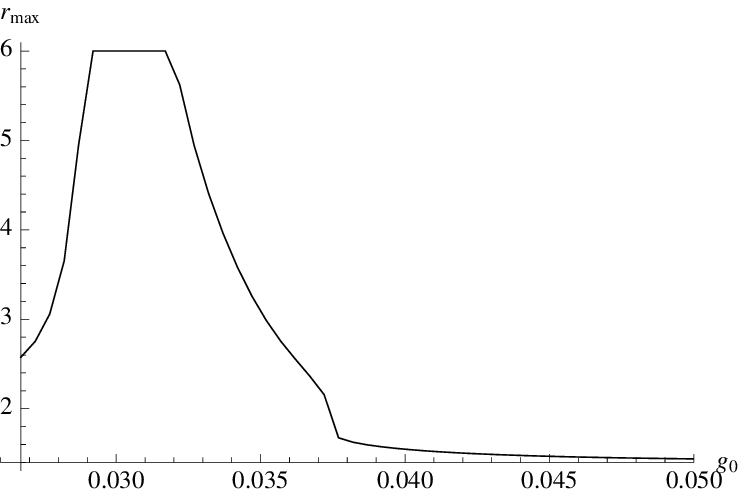}
\includegraphics[width=0.48\textwidth]{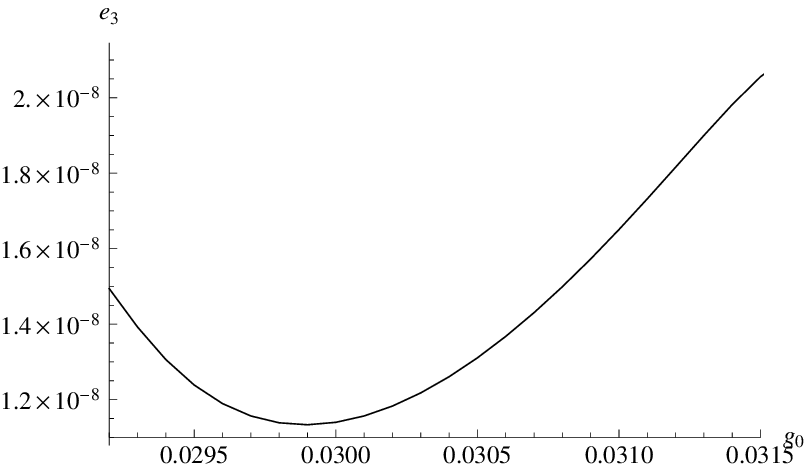}
\caption{\label{fig.regconmin}
Summary of the results obtained from the numerical shooting algorithm. Solutions satisfying the boundary condition $e_2$, characterized by the two free parameters $(g_0, g_1)$ are shown in the top diagram. Points with $e_2(g_0, g_1)>0$ are marked green while points with $e_2(g_0, g_1)<0$ are marked red. The black line indicates the parameters which lead to solutions which are regular at $r_{1, {\rm sing}}$ and $r_{2, {\rm sing}}$. Along this line $g_1$ is determined as a function of $g_0$. The extension of these solutions to the interval $I_2$ is depicted in the bottom diagrams. Notably, most of the solution do not extend to $r_{3, {\rm sing}}$ but terminate in a movable singularity. The point of termination $r_{\rm max}$ of the numerical solution as a function of $g_0$ is shown in the left panel. For the solutions extending to $r_{3, {\rm sing}}$, the regularity condition $e_3(g_0(g_1))$ is evaluated in the right panel. The minimum shown by $e_3(g_0)$ indicates that there is one isolated solution which satisfies all three BCs, $e_i = 0$.
}
\end{center}
\end{figure}
The values for $e_2$ obtained in this way are shown in the top diagram
of Fig.\ \ref{fig.regconmin}. Here lattice points with $e_2 > 0$ are marked
green while points with $e_2 < 0$ are indicated in red. The regular points
satisfying $e_2 \approx 0$ are given by the black line. This
 set of regular points constitutes a one dimensional subspace that will be referred to as regular line.
Every point on the regular line can be identified by $g_0$, which is the only remaining free parameter.
Thus implementing regularity of the solution at the fixed singularities $r_{1, {\rm sing}}$ and $r_{2, {\rm sing}}$ reduces the dimension of space of local solutions by two.

In the next step, we cross the fixed singularity by utilizing that the solutions regular at $r = r_{2, {\rm sing}}$ admit the polynomial expansion \eqref{eq:expansions}. 
 The two coefficients $b_0$ and $b_1$ are determined by matching the expansion with the numerical solution \eqref{numsol1}
\eq{
	\label{matching}
	\varphi(r_2 - \epsilon_2;g_0) = \varphi_2(r_2 - \epsilon_2;b_0,b_1),\qquad
	\varphi'(r_2 - \epsilon_2;g_0) = \varphi'_2(r_2 - \epsilon_2;b_0,b_1)\,.
}
Here we explicitly indicated that the numerical solution depends on the single remaining free parameter $g_0$, stressing that we consider the regular line only.
Given the map
\be
g_0 \mapsto ( b_0, b_1 )
\ee
we evaluate the polynomial expansion at $r_{2, {\rm sing}} + \epsilon_2$ in order to obtain the initial conditions
for the numerical extension of the regular solutions to the interval $I_2$ 
\spliteq{\label{eq:inits2}
\vp(r_{2} + \epsilon_2)	&=  \varphi_2(r_{2} + \epsilon_2; g_0)\,, \\
\quad
\vp'(r_{2} + \epsilon_2) &	=  \varphi_2'(r_{2} + \epsilon_2; g_0)\,, \\
\quad
\vp''(r_{2} + \epsilon_2)	& =  \varphi_2''(r_{2} + \epsilon_2; g_0)\,.
}

The results obtained from the second numerical integration are shown in the bottom diagrams of Fig.\ \ref{fig.regconmin}.
Notably, only a small window of points on the regular line, $ 0.029 \lesssim g_0 \lesssim 0.0315$,
the solutions can be extended up to $r_{3, {\rm sing}} - \epsilon_3$. All other solutions
terminate in a movable singularity. Their corresponding end points are displayed
in the bottom-left diagram of Fig.\ \ref{fig.regconmin}. For the solutions reaching
$r_{3, {\rm sing}} - \epsilon_3$ the numerical integration provides
a map
\eq{
g_0 \mapsto e_3(r_3-\epsilon_3;g_0) \, .
}
The values of $e_3(r_3-\epsilon_3;g_0)$ are shown in the bottom-right diagram of Fig.\ \ref{fig.regconmin}.
The map has a pronounced minimum at $g_0 \approx 0.0299$. At this minimum
the BC $e_3 \approx 0$ is satisfied within the numerical accuracy of the shooting algorithm. 
This indicates that there is a single, isolated solution which is also regular at $r_{3, {\rm sing}}$.

\begin{figure}[t]
	\begin{center}
		\includegraphics[width=0.95\textwidth]{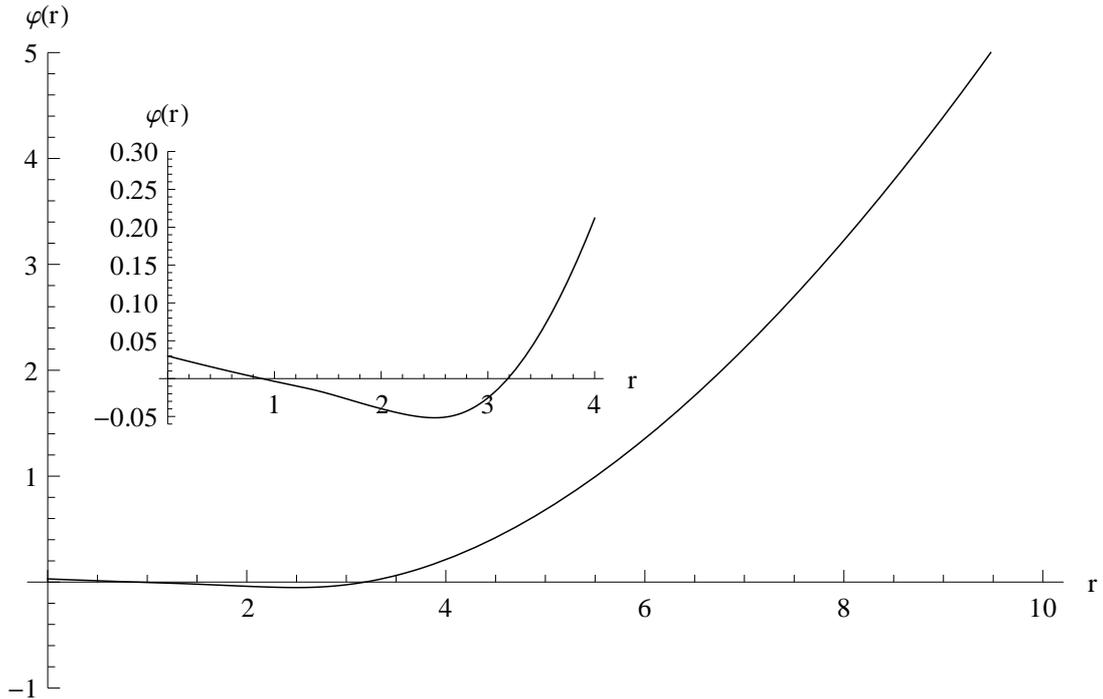}
		\caption{\label{fig:numsol}
			The isolated, globally well-defined fixed function $\varphi_*(r)$ constructed from imposing the three regularity conditions \eqref{eq:BCs}. The fixed function is bounded from below, comes with a  positive cosmological constant $\lambda_*$ and Newton's constant $g_*$, and displays a global minimum at $r_{\min} \approx 2.5$.
		}
	\end{center}
\end{figure}

\begin{table}[t]
	\centering
	\renewcommand\arraystretch{1.5}
	\begin{tabular}{|l|c|c||c||c|c|c|c|}
		\hline
	 &	$g_0$ & $g_1$ & $\tau_*$ & $b_0$ & $b_1$& $c_0$ & $c_1$ \\ \hline \hline
	full flow 	& $0.0299$ & $-0.0364$ & $0.224$ & $-0.0154$ & $-0.0342$& $1.3266$ & $0.7521$\\ \hline \hline
	conformally reduced & $0.0528$ & $-0.0587$ & $0.152$ & $-0.0198$ & $-0.0448$& $13.558$ & $7.1777$\\\hline
	polynomial expansion & $0.0111$ & $-0.0237$ & $0.197$ & $-$ & $-$ & $-$ & $-$ \\ \hline \hline
	\end{tabular}
	\caption{Numerical values of the first two coefficients appearing in the polynomial expansions \eqref{eq:gexp}, \eqref{eq:expansions}, and \eqref{eq:cexp} of the fixed function $\varphi_*$
at the fixed singularities \eqref{eq:poleposition}. The first line shows the coefficients obtained from the full flow equation. The values from the conformally reduced ODE \eqref{eq:fpeqCR} are derived in Appendix \ref{App:A} and shown in the second line. For completeness, we also give the expansion coefficients of the polynomial $f(R)$-expansion obtained in \cite{Demmel:2014hla} in the third line. The universal product $\tau_* \equiv g_0/(32 \pi g_1^2)$ is qualitatively similar in all three cases.}
	\label{tab:coefficients}
\end{table}
Finally, the isolated solution is extended to the interval $I_3$. Following
the strategy employed when crossing the singularity at $r_{2, {\rm sing}}$,
the numerical solution is matched to the polynomial expansion \eqref{eq:cexp}
\eq{
	\varphi(r_3 - \epsilon_3;g_0) = \varphi_3(r_3 - \epsilon_3;c_0,c_1) \, , \quad
	\varphi'(r_3 - \epsilon_3;g_0) = \varphi'_3(r_3 - \epsilon_3;c_0,c_1)\,.
}
This determines the values of $c_0$ and $c_1$ for the regular solution.
Following the strategy implemented in eq.\ \eqref{eq:inits2},
 the parameters $c_0, c_1$ are converted into initial conditions for a numerical integration on $I_3$. 
The solution $\varphi_*(r)$ is then completed by carrying out this numerical integration
and matching to the 
scaling regimes
discussed in Sect.\ \ref{sect:4.2}.\footnote{Based on the values $c_0, c_1$, the the solution
 on the interval $I_3$ can also be obtained by recursively
 solving the polynomial expansion \eqref{eq:cexp}. Carrying out the expansion
 up to $N=10$, one finds that this expansion agrees with the numerical solution within its radius of convergence.}
The numerical values of all expansion coefficients for the regular solution are summarized in the first line of Tab.\ \ref{tab:coefficients}.
The resulting isolated and globally well-defined fixed function $\varphi_*(r)$
is shown in Fig.\ \ref{fig:numsol} and its properties are discussed further in Sect.\ \ref{sect:5.3}.
This solution is the first proper fixed function obtained within four-dimensional Quantum Einstein Gravity and constitutes the central result of this work.

We close this subsection with some technical comments on the implementation of the shooting algorithm.
Notably, the position and width of the plateau displayed in the bottom-left diagram of Fig.\ \ref{fig.regconmin}
has a mild dependence on the parameters entering the numerical integration.
It is conceivable that it reduces to a single point (spike) when implementing a higher numerical precision.
Owing to the fact that we are using a polynomial expansion at the singular points to continue 
the solutions through a fixed singularity the entire construction is very robust 
under variations of the auxiliary parameters $\epsilon_1$, $\epsilon_2$ and $\epsilon_3$.
Changes in $\epsilon_i$ induce small changes (within error-bars) of $g_0$
that propagate into the coefficients $b_0, b_1$ and $c_0, c_1$.
These quantitative changes do not affect the qualitative
picture shown in Fig.\ \ref{fig.regconmin}, however.

\subsection{Properties of the complete fixed function}
\label{sect:5.3}
The proper fixed function $\varphi_*(r)$ satisfying all BCs \eqref{eq:BCs} is shown in Fig.\ \ref{fig:numsol}.
Here, we will summarize its main properties. \\

\noindent
{\bf Asymptotic expansion in the UV} \\
At $r = 0$, which corresponds to $k \rightarrow \infty$ if the background curvature is held fixed, the fixed function admits the analytic expansion \eqref{eq:gexp}. The first two expansion coefficients are given in the first line of Tab.\ \ref{tab:coefficients}. In order to facilitate the comparison with results obtained in finite-dimensional approximations, it is convenient to compute the corresponding values of the dimensionless Newton's constant $g_* \equiv -(16 \pi g_1)^{-1}$ and cosmological constant $\lambda_* \equiv -g_0/(2 g_1)$ together with the universal (gauge-independent) product $\tau_* = g_* \, \lambda_*$
\be
\lambda_* = 0.411 \, , \qquad g_* = 0.547 \, , \qquad \tau_* = 0.224 \, . 
\ee
These values are in good agreement with the results obtained from
 expansions of flow equations truncated at finite order $R^N$ \cite{Reuter:1996cp,Lauscher:2001ya,Reuter:2001ag,Litim:2003vp}. In
particular $\tau_*$ fits well to results obtained within ``geometric'' flow equations \cite{Donkin:2012ud,Demmel:2014hla}.
 We take this agreement as strong evidence that, despite the different boundary conditions imposed on the flow equations,
 the polynomial expansions and the fixed function $\varphi_*(r)$ actually describe the same universality class. \\
 
\noindent
{\bf Consequences of the global minimum} \\
The fixed function possesses a global minimum located at $r_{\min} \approx 2.500$. 
At this minimum $\varphi_*(r_{\min}) < 0$. This feature together with
the resulting non-monotonicity distinguishes our solution from earlier constructions \cite{Benedetti:2012dx,Dietz:2012ic},
where $\varphi_*(r)$ turned out to be positive definite and monotonically increasing.

As a consequence our solution actually \emph{passes the redundancy test} \cite{Dietz:2013sba},
requiring that
\be
E_4(r) \equiv 2 \varphi_*(r) - r \, \varphi_*'(r)  
\ee
has a zero for some value $r$. From Fig.\ \ref{fig:numsol} we then deduce
\be
E_4(0) = 2 \, g_0 > 0 \, , \qquad E_4(r_{\rm min}) = 2 \, \varphi_*(r_{\min}) < 0 \, . 
\ee
Since $E_4(r)$ is continuous this implies that $E_4(r)$ has a zero within the interval $r \in [0, r_{\rm min}]$.
This provides a strong indication that linear perturbations of the fixed function cannot be absorbed by a 
field redefinition and constitute genuine eigenperturbations of the fixed function.\\

\begin{figure}[t]
	\begin{center}
		\includegraphics[width=0.6\textwidth]{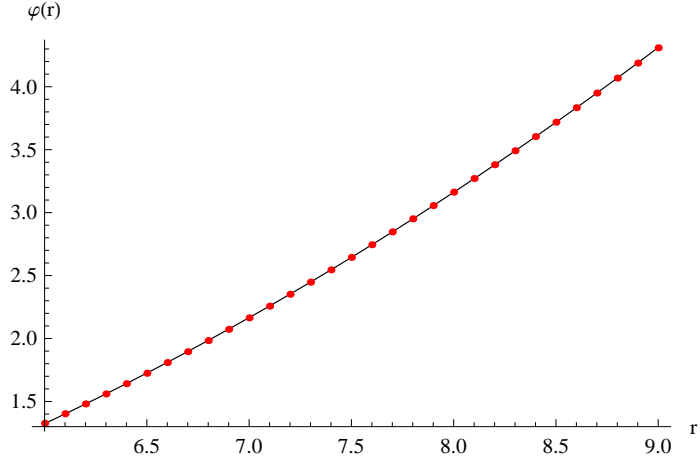}
		\caption{\label{fig:asymptoticfit}
		The numerical solution (red points) and the non-linear fit of the ansatz \eqref{eq:fitansatz}. The values of the fit parameters are given in eq.\  \eqref{eq:fitparamters}.
		}
	\end{center}
\end{figure}

\noindent
{\bf Asymptotic behavior in the IR} \\
Finally, it is interesting to investigate the asymptotic behavior of $\varphi_*(r)$ for $r \rightarrow \infty$.
In order to determine the scaling regime, we first approximate the large-$r$ behavior through
the ansatz
\eq{\label{eq:fitansatz}
\vphi(r) = A \, r^2 \left(1 + u_1 \, r^{-1} + u_2 \, r^{-2} \right) \, , 
}
neglecting further, sub-leading terms. The leading $r^2$ behavior
is motivated by the classical and balanced scaling asymptotics encountered in Sect.\ \ref{sect:4.2}.
Fitting eq.\ \eqref{eq:fitansatz} to the numerical solution obtained on the interval $r \in [6,9]$
yields the parameters
\be\label{eq:fitparamters}
\begin{split}
A^{\rm fit} = & \, 0.07705 \pm 0.00032 \, , \\
u_1^{\rm fit} = & \, -2.07514 \pm 0.05399 \, , \\
u_2^{\rm fit} = & \, -6.36855 \pm 0.25897 \, . 
\end{split}
\ee
Fig.\ \ref{fig:asymptoticfit} shows that
the ansatz indeed provides a good approximation of the numerical solution in this regime.

At this stage it is also illuminating to substitute the ansatz \eqref{eq:fitansatz} into the ODE \eqref{eq:fpeq} 
and determine the coefficients $A$, $u_1$ and $u_2$ from a large-$r$ expansion. Surprisingly,
the ansatz provides an asymptotic solution if and only if the endomorphisms satisfy
\be
43 + 300 \alpha + 360 \alpha^2 - 36 \beta + 108 \beta^2 = 0 \, .
\ee
This condition is not satisfied for the choice \eqref{eq:endochoice} and becomes
compatible with the ELE condition \eqref{ELE} for two specific values $\beta_\pm = (18 \pm \sqrt{285})/78$ only.
Therefore we conclude that the asymptotic behavior of $\varphi_*(r)$ 
is described by the scaling solution obtained in the balanced regime
and involves also non-analytic terms not present in the ansatz \eqref{eq:fitansatz}.
This is actually in agreement with earlier
works \cite{Dietz:2012ic,Bonanno:2012jy,Dietz:2013sba} which also observed the presence of non-analytic terms in this
asymptotic scaling regime.

\section{Conclusions and discussion}
\label{sect.6}
The existence of a NGFP on the gravitational theory space is a crucial ingredient for constructing a consistent and predictive quantum theory of gravity along the lines of Weinberg's Asymptotic Safety scenario. In this work, we used the functional renormalization group equation for the effective average action $\Gamma_k$, \eqref{FRGE}, to establish the existence of a suitable fixed function in the realm of $f(R)$-gravity in four spacetime dimensions.
Despite the vast body of earlier works \cite{Benedetti:2012dx,Benedetti:2013nya,Dietz:2012ic,Dietz:2013sba}, this is the first time that a suitable fixed function has been found and its construction constitutes an important step towards generalizing finite-dimensional truncations of the gravitational RG flow to approximations containing an infinite number of scale-dependent coupling constants.
Our solution is globally well-defined on the positive real axis and its UV-expansion at $r = 0$ yields a positive cosmological constant and Newton?s constant. The structure and properties of the fixed function is in good agreement with the fixed points solutions found at the level of polynomial $f(R)$-truncations \cite{Demmel:2014hla}. This construction constitutes an important step towards generalizing finite-dimensional truncations of the gravitational RG flow to approximations containing an infinite number of scale-dependent coupling constants.

The fixed function is based on non-linear ODE of third order, which construction principles discussed in the introduction.
Owing to the inclusion of non-trivial endomorphisms in the regulator the ODE possesses three fixed singularities at finite $r \equiv R/k^2$.
This feature restricts the space of globally well-defined solutions to one isolated fixed function. The condition of ``equal lowest eigenvalues'' thereby plays an important role for the existence of this solution: numerical analysis indicates that once the transverse-traceless and scalar sectors of the flow equation are out of tune, the numerical solutions typically terminate in a movable singularity.
Combining a scaling analysis of the ODE together with the numerical solution shows that $f_*(R) = A R^2 (1 + \eta \log R + \dots) $, matching the IR behavior of earlier constructions 
\cite{Benedetti:2012dx,Benedetti:2013nya,Dietz:2012ic,Dietz:2013sba}.

In Appendix \ref{App:A} we analyzed conformally reduced version of the full flow equation. The resulting ODE shares many properties with the full flow. In particular it is of the same order and exhibits the same fixed singularities as the full system. At the same time the conformally reduced system is significantly simpler which allows investigating the existence of fixed functions by employing complementary techniques. As the central result we established that the conformally reduced equation also exhibits one globally well-defined isolated fixed function whose properties are qualitatively similar to the ones found in the full system (cf.\ Figs.\ \ref{fig:numsol} and \ref{fig:numsolCR}). The result establishes that conformally reduced approximation is a very useful setting when searching for fixed functions of the gravitational flow equation.

At this stage, we have not yet analyzed the structure of admissible, globally well-defined, infinitesimal deformations of the fixed function. Substituting the corresponding ansatz describing these deformations into the PDE \eqref{eq:smoothapprox} leads to a linear third-order ODE which also contains the (possibly complex) scaling exponent of the deformation. In principle the corresponding free parameters can again be determined by applying the combination of analytical and numerical techniques employed in the construction of the fixed function. This numerically very demanding, challenge is, however, beyond the scope of the present work. The match between the fixed function and its finite-dimensional sibling let us expect that the full functional also shares the stability properties of the polynomial approximation. Presupposing that all relevant eigenvalues can be extended to global deformations, this implies that there should be three relevant deformations.
This expectation that there is a finite number of relevant directions is also supported by the general arguments \cite{Benedetti:2013jk}.
As an important first check in this direction, we explicitly verified that our solution passes the redundancy test \cite{,Dietz:2013sba}, indicating that the fixed functional might indeed have non-trivial deformations which cannot be absorbed by a field redefinition.

Let us again stress that carrying out the extension of the fixed points seen in finite-dimensional truncations to the level of fixed functions is an important step in the Asymptotic Safety program: the functional approximations are much more sensitive to structural details of the theory, as, e.g., the functional measure used to construct the flow equations. These details have a crucial influence on the singularity structure of the ODE describing the fixed function and may provide important insights on conceptual properties required to achieve Asymptotic Safety. In this context, we find it encouraging that the geometrically motivated flow equation actually admits an interesting fixed function solution.

Notably, our findings are not in tension with the recent results \cite{Dietz:2015owa} where no fixed function for four-dimensionally conformally reduced gravity has been found. The flow equation and search strategy implemented in this work differs from our setting in various ways, which can be made explicit based on the construction principles detailed in the introduction. Firstly, the flow equation constructed in \cite{Dietz:2015owa} utilizes a conformally flat background while our equation has been derived for a compact background with positive curvature. Secondly, the analysis in \cite{Dietz:2015owa} implemented the modified split ward identities of the effective average action to reduce the fixed point equation to a linear PDE. While this certainly simplifies the corresponding fixed function analysis the special choice of regulator or the implementation of split symmetry restoration at the level of the truncation may be overconstraining the system such that no globally well-defined solution may be found.

In this light, it would be very interesting to investigate to which extend the construction principles of our proposal are strictly necessary for the existence of a satisfactory fixed function.
In particular it would be interesting to investigate if a flow equation based on the exponential parametrization of the metric fluctuations
\be
g_{\mu\nu} = \gb_{\mu\rho} \, \left( e^h \right)^\rho{}_\nu \, ,
\ee
recently advocated in \cite{Nink:2014yya,Percacci:2015wwa,Falls:2015qga}
also gives rise to fixed functions at the level of the $f(R)$-approximation.
Compared to previous constructions, this parameterization corresponds to a change in the path integral measure. Combined with a clever choice of gauge-fixing this may lead to a second order ODE describing the fixed function. This structure is much simpler to analyze since fixed functions may be identified based on a single shooting instead of the double-shooting method required in this work. In \cite{Percacci:2015wwa,Falls:2015qga} it was already observed that this choice significantly improves the IR behavior of the gravitational RG flow.

\bigskip

\noindent
{\it Acknowledgements} We thank M.\ Reuter, A.\ Bonanno, T.R.\ Morris and K.\ Falls for interesting discussions. The work by F.S.\
is supported by FOM grants 13PR3137 and 13VP12.

\begin{appendix}
\section{Fixed functions of the conformally reduced flow equation}
\label{App:A}
In this appendix, we investigate the fixed functions entailed by the conformally reduced flow
equation 
\eqref{eq:fpeqCR}. We establish that the conformally
reduced setting gives rise to one fixed function whose properties are 
qualitatively very similar to the one constructed from the full flow equation. The main advantage of the conformally reduced system is that it is structurally significantly simpler, so that the structure and properties of the resulting solution $\varphi_*(r)$ can be corroborated using both analytical and numerical methods. Following
our choice of endomorphisms from the main part of the paper, we will work with $\beta = \frac{1}{6}$ throughout this section.

\subsection{Fixed functions from the shooting method}

We start by analyzing the singularity structure of \eqref{eq:fpeqCR}. Since the positions of the fixed singularities are determined by the scalar trace, the discussion of Sect.\ \ref{sect.4} can be applied to \eqref{eq:fpeqCR} and the the fixed singularities are again given by \eqref{eq:poleposition}.
The boundary conditions that fixed functions have to satisfy are determined by Laurent expanding according to \eqref{eq:laurentexp} and explicitly read
\spliteq{\label{eq:BCsCR}
	e_1
	&=
	\frac{7 \varphi ''}{2}+\frac{7 \varphi '}{9}-\frac{256}{9} \pi ^2 \varphi \left(9 \varphi ''+3 \varphi '+2 \varphi \right)
	\,,
	\\
	e_2
	&=
-\frac{96}{65} \varphi ''-\frac{97}{90} \varphi '+\frac{256 \pi ^2 \left(13 \varphi
	-9 \varphi '\right) \left(72 \varphi ''-39 \varphi
	'+169 \varphi \right)}{7605}
\,,
\\
e_3
&=
9 \left(\varphi '-12 \varphi ''\right)-512 \pi ^2 \left(\varphi -3 \varphi '\right) \left(18 \varphi ''-6 \varphi '+\varphi \right)
\,.
}

Any regular solution satisfying the BCs \eqref{eq:BCsCR} can be expanded in a Taylor series at $r_{i, {\rm sing}}$ according to \eqref{eq:gexp}, \eqref{eq:expansions} and \eqref{eq:cexp}.
Substituting these expansions into \eqref{eq:fpeqCR} allows for determining all coefficients $g_n$, $b_n$ and $c_n$ with $n\geq2$ in terms of the first two coefficients $(g_0,g_1)$, $(b_0,b_1)$ and $(c_0,c_1)$, respectively.

\begin{figure}[t]
	\begin{center}
		\includegraphics[width=0.7\columnwidth]{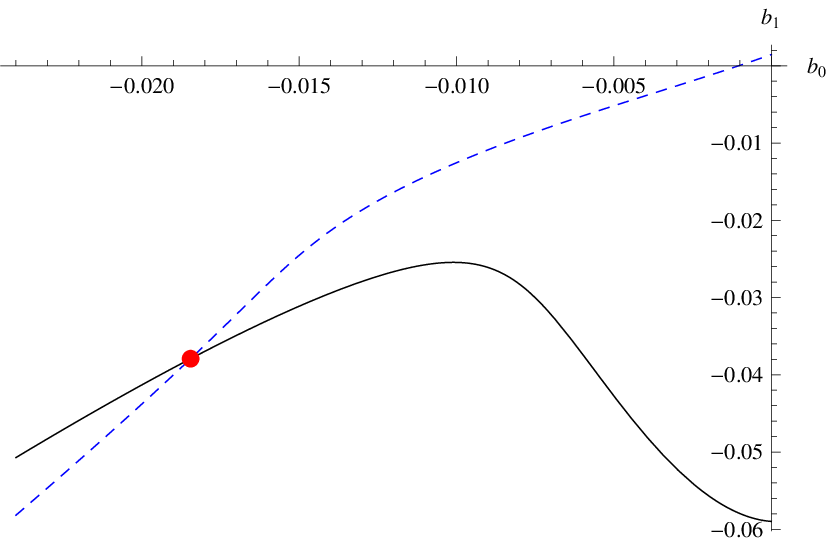}
		\caption{\label{fig:regconmin2CR}
			The black line shows $b_1$ as a function of $b_0$ obtained by imposing the bootstrap condition \eqref{simplified2}, i.e.\ $b_{M+1}(b_0,b_1)=0$, for $M=3$.
			Due to the asymptotic nature of the polynomial expansion \eqref{eq:expansions} at $r=r_{2,{\rm sing}}$, it is impossible to go to higher orders.
			The blue line is the result of the bootstrap condition $g_{N+1}(g_0,g_1)=0$ for $N=17$ transported to $r_{2, {\rm sing}}$ by using the Taylor expansion.
			The two lines intersect at $b_0 \approx -0.0185$ and $b_1 \approx -0.0379$ corresponding to $g_0 \approx 0.0444$ and $g_1 \approx -0.0512$.
			The bootstrapping strategy confirms the numerical results and thus, the existence of globally well-defined fixed functions for the conformally reduced system.
		}
	\end{center}
\end{figure}

In order to construct globally well-defined solutions of \eqref{eq:fpeqCR} we repeat the numerical shooting algorithm used in Sect.\ \ref{doubleshooting}.
Beginning at $r=0$ we have a two-parameter family of solutions characterized by $(g_0,g_1)$.
To extend these solutions to $[0,\infty[$, we again split the positive half-axis into the intervals \eqref{intervals}. 
The first shooting on $I_1$ is realized by numerically integrating from $r=\epsilon_1$ to $r_{2,{\rm sing}} - \epsilon_2$ using the initial values \eqref{eq:inits1}.
After discretizing the $(g_0,g_1)$-plane we perform for every lattice point the integration and evaluate the BC $e_2$ with it.
This provides a map
\eq{\label{CRmap}
(g_0,g_1) \mapsto e_2(r_2 - \epsilon_2;g_0,g_1)\,.
}
The resulting values $e_2$ are visualized in Fig.\ \ref{fig:regconmin2CR}, where lattice points with $e_2>0$ are marked green while points with $e_2<0$ are tagged red.
Regular points have to satisfy $e_2 \approx 0$.
Clearly, there is a line of points separating two regions of different signs (i.e.\ different colors in the diagram) giving rise to a line of regular solutions. The points on
this line are again parametrized by $g_0$.

\begin{figure}[t]
	\begin{center}
		\includegraphics[width=0.7\columnwidth]{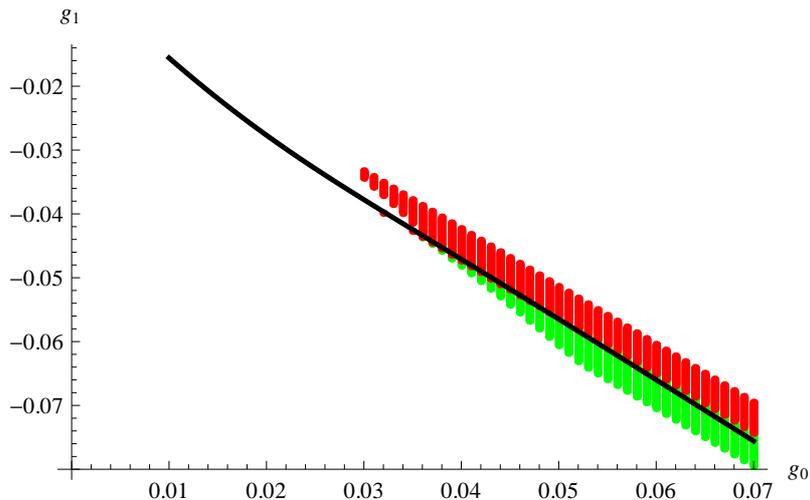}
		\caption{\label{fig.regconminCR}
			Solutions regular at $r=0$ are parametrized by $g_0$ and $g_1$. Evaluating these solutions at the BC $e_2$, eq.\ \eqref{eq:BCsCR}, yields the data points depicted above.
			Points with $e_2>0$ are marked green while points with $e_2<0$ are tagged red.
			There is a line of points separating two regions of different signs (i.e.\ different colors) giving rise to a line of regular solutions satisfying $e_2\approx 0$.
			The black line shows $g_1$ as a function of $g_0$ obtained by imposing the bootstrap condition $g_{N+1}(g_0,g_1)=0$ for $N=17$ and solving for $g_1$.
			The two approaches show very good agreement.
		}
	\end{center}
\end{figure}

\begin{figure}[t]
	\begin{center}
		\includegraphics[width=0.9\columnwidth]{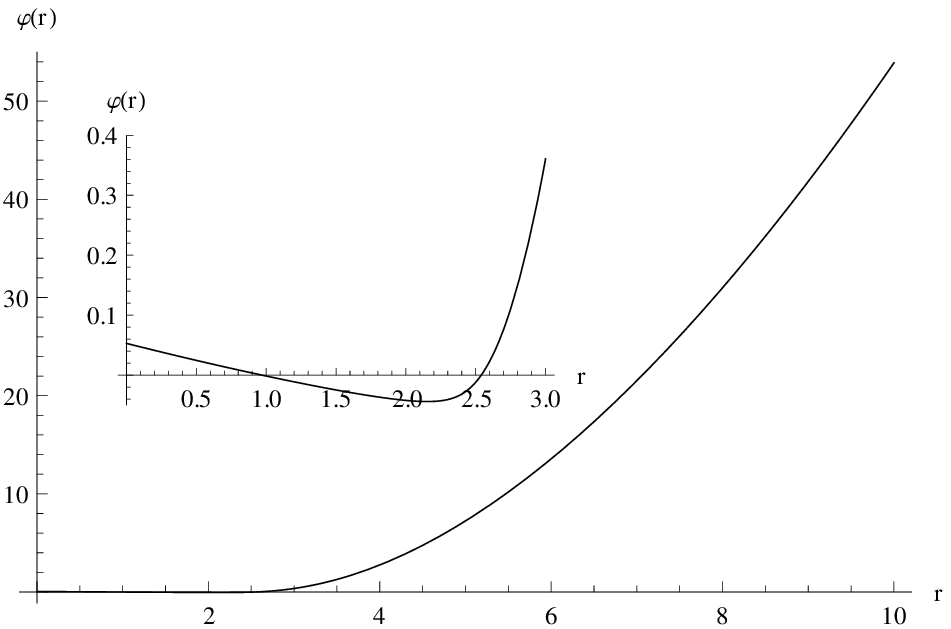}
		\caption{\label{fig:numsolCR}
			The isolated, globally well-defined fixed function $\varphi_*(r)$ of the conformally reduced system constructed from imposing the three BCs \eqref{eq:BCsCR}. The fixed function is bounded from below, comes with a  positive cosmological constant $\lambda_*$ and Newton's constant $g_*$, and displays a global minimum at $r_{\min} \approx 2.153$.
		}
	\end{center}
\end{figure}

Next, we repeat the shooting procedure on $I_2$.
Utilizing the polynomial expansion at $r_{2,{\rm sing}}$, initial values for the numerical integration are again determined by matching according to \eqref{matching}.
Integrating from $r_{2,{\rm sing}} +\epsilon_2$ to $r_{3,{\rm sing}}-\epsilon_3$ again reveals a window of $g_0$-values where the
corresponding solutions extend up to $r_{3,{\rm sing}}$. 
This regular window is qualitatively similar to the one displayed in Fig.\ \ref{fig.regconmin} and located at $ 0.052 \lesssim g_0 \lesssim 0.054$.
The unique value for $g_0$ minimizing the last BC $e_3$ is listed in Tab.\ \ref{tab:coefficients} and the corresponding globally well-defined fixed function $\vphi_*(r)$ is displayed in Fig.\ \ref{fig:numsolCR}.
It shares all features of the result obtained from the full flow \eqref{eq:fpeq}, including a minimum located at $r_{\rm min} \approx 2.153$.
This result underlines the importance of the conformal mode for the existence of fixed functions.

\subsection{Fixed functions from bootstrapping method}

We close this section by presenting an alternative bootstrapping strategy relying on polynomial expansions only \cite{Abbasbandy:2011ij}.
In the conformally reduced setup all expansion coefficients $g_n$, $b_n$ and $c_n$ are much smaller in size enabling us to go to much higher orders in the Taylor expansions, especially at $r=0$.
Thus, we start by truncating the Taylor series \eqref{eq:gexp} to a finite polynomial of the order $N$
\eq{
\label{finitegexp}
\varphi_1(r;g_0,g_1) = g_0 + g_1 r + \sum_{n=2}^{N} g_n(g_0,g_1) \, r^n\, . 
}
If the Taylor series \eqref{eq:gexp} has a finite radius of convergence $R_1$ and if additionally $r_{2,{\rm sing}} \leq R_1$, then \eqref{finitegexp} can be substituted into the BC $e_2$.
Again, this provides a map of the form \eqref{CRmap}.
The condition $e_2(g_0,g_1)=0$ relates $g_0$ and $g_1$ analytically.
Unfortunately, this relation grows quickly in complexity with increasing order $N$ and thus, is of no real practical use.
However, if the finite Taylor polynomial \eqref{finitegexp} serves a good approximation at the order $N$, the next coefficient should by very small, i.e.\ $g_{N+1}(g_0,g_1) \approx 0 $.
Thus, a much simpler condition that can be imposed is given by
\eq{
\label{simplified}
g_{N+1}(g_0,g_1)=0\,,
}
which constitutes the central idea of the bootstrapping strategy (or ``minimal procedure'' in the terminology of \cite{Abbasbandy:2011ij}).
Being a rational function in $g_0$ and $g_1$, at every order new solutions of \eqref{simplified} emerge, which are mostly complex or singular and thus, discarded as spurious.

Pushing the order up to $N=17$, only one real and non-singular solution of $g_{18}=0$ has been found and is displayed in Fig.\ \ref{fig.regconminCR} (black line).
Most remarkably, the analytic solution $g_0 \mapsto g_1$ is in perfect agreement with the regular line obtained from the shooting algorithm.
Moreover, no information from the second pole has been incorporated yet.
This can be done by repeating the bootstrap strategy at $r_{2,{\rm sing}}$ and imposing
\eq{
\label{simplified2}
	b_{M+1}(b_0,b_1) = 0\,,
}
where $M$ is the order to which the expansion \eqref{eq:expansions} has been truncated.
Imposing \eqref{simplified2} yields a line $b_0 \mapsto b_1$ in the $(b_0,b_1)$-plane (spurious solutions are discarded), c.f.\ Fig.\ \ref{fig:regconmin2CR} (black line).
In contrast to the expansion at $r=0$, no convergence of the line can be observed revealing the asymptotic nature of the expansion at $r_{2,{\rm sing}}$.
Thus, we keep the order $M$ low ($M=3$ at most) and ``transport'' the previously obtained regular line from $r=0$ to $r_{2,{\rm sing}}$ into the $(b_0,b_1)$-plane.
This is done by substituting the analytic solution $g_0 \mapsto g_1$ into the Taylor polynomial \eqref{finitegexp} and simultaneously setting $r=r_{2,{\rm sing}}$.
The transported line is defined as
\eq{
g_0 \mapsto (\vphi_1(r_{2,{\rm sing}};g_0,g_1(g_0),\vphi'_1(r_{2,{\rm sing}};g_0,g_1(g_0)) \equiv (b_0(g_0),b_1(g_0))\,,
}
and shown (blue line) in Fig.\ \ref{fig:regconmin2CR}.
Inspecting Fig.\ \ref{fig:regconmin2CR} shows that the transported line (blue line) and the solution of \eqref{simplified2} (black line) intersect at
\eq{
	\label{firstmatch}
	b_0 \approx -0.0185 \, \qquad b_1 \approx -0.0379\, ,
}
corresponding to
\eq{
	g_0 \approx 0.0444, \qquad g_1 \approx -0.0512\,.
}
Comparing with the values from the shooting method in Tab.\ \ref{tab:coefficients} shows a good qualitative agreement.
In summary, the bootstrap strategy provides a further argument in favor of the existence of a globally well-defined fixed function.
However, due to the asymptotic nature of the expansion \eqref{eq:expansions} a precise determination of the coefficients characterizing the fixed function is impossible.
Thus the shooting method employed in the main part of this work constitutes the most reliable method for constructing fixed functions in this setting.

\end{appendix}


\end{document}